\documentclass[fleqn,usenatbib]{mnras}

\usepackage{newtxtext,newtxmath}

\usepackage[T1]{fontenc}
\usepackage{amsmath}
\usepackage{graphicx}

\usepackage{lipsum}
\usepackage{hyperref}
\usepackage{mathtools, nccmath}
\usepackage{longtable}
\usepackage{lineno}
\usepackage{orcidlink}
\usepackage{enumitem}
\usepackage{amsmath}
\usepackage{amsfonts}

\DeclareRobustCommand{\VAN}[3]{#2}
\let\VANthebibliography\thebibliography
\def\thebibliography{\DeclareRobustCommand{\VAN}[3]{##3}\VANthebibliography}

\usepackage{graphicx}	
\usepackage{amsmath}	
\usepackage{threeparttable}

\title[Neural Networks for AMI deconvolution]{Revealing Io's Surface using JWST-NIRISS Aperture Masking Interferometry and Neural Network Deconvolution}

\author[J. Sanchez-Bermudez et al.]{
J. Sanchez-Bermudez,$^{1}$ \orcidlink{0000-0002-9723-0421}  \thanks{E-mail: joelsb@astro.unam.mx}
I. de Pater,$^{2}$
A. Conrad$^{2,3}$
A. Sivaramakrishnan$^{4}$,
E. Molter$^{4}$,
D. Thatte$^{4}$,
\newauthor R. Cooper$^{4}$,
K. de Kleer$^{5}$,
L. Roth$^{6}$
\\
$^{1}$Universidad Nacional Aut\'onoma de M\'exico. Instituto de Astronom\'ia. A.P. 70-264, 04510, Ciudad de M\'exico, 04510, M\'exico\\
$^{2}$University of
California—Berkeley, Berkeley, CA, USA\\
$^{3}$Large Binocular Telescope Observatory, The University of Arizona, Tucson, AZ, USA\\
$^{4}$Space Telescope Science Institute, 3700 San Martin Drive, Baltimore, MD 21218, USA\\
$^{5}$California Institute of Technology, Pasadena, CA 91125 USA\\
$^{6}$Royal Institute of Technology, Stockholm 104 50, Sweden
}

\date{Accepted XXX. Received YYY; in original form ZZZ}

\pubyear{\the\year{}}

\begin{document}
\label{firstpage}
\pagerange{\pageref{firstpage}--\pageref{lastpage}}
\maketitle

\begin{abstract}

Io is one of Jupiter's largest moons and the most volcanically active body in the Solar System. Its very active surface has hot spots produced by volcanic eruptions popping up at seemingly random locations and times. Characterizing the complex surface of Io requires the highest angular resolution available. This work presents the analysis of Aperture Masking Interferometric observations (at 4.3 $\mu$m) of Io taken with the Near-Infrared Imager and Slitless Spectrograph instrument on the James Webb Space Telescope. These are the first space-based infrared interferometric observations of a Solar System body ever taken.  For complex extended objects like Io, the traditional visibility extraction algorithms from interferograms suffer from limitations. Here, new deconvolution methods based on Neural Networks allowed us to obtain reliable images from which a detailed analysis of the volcanically active surface of this moon was performed. Our study characterizes the loci and brightness of several unresolved volcanoes on the surface of Io, as well as the extended emission observed. We identified the brightest eruption (I$_{\mathrm{4.3 \mu m}}$ = 33 $\pm$ 4.3 GW/ $\mu$m), referred to as V1, within an area to the North-East of Seth Patera (129.4 $\pm$ 0.8$^{\circ}$ W. Longitude, 1.5 $\pm$ 0.7$^{\circ}$ S. Latitude). Its projected speed (V$_{\mathrm{T}}$ = 86 $\pm$ 34 m s$^{-1}$) is consistent with the rotational speed of Io. Additionally, six fainter volcanoes were identified and characterized. Complementary ground-based images, taken with the Keck II telescope, allowed us to benchmark the deconvolved Aperture Masking Interferometric images, showing consistency. Finally, we highlight the importance of characterizing Io's surface with long-term monitoring at high-angular resolution.  

\end{abstract}

\begin{keywords}
techniques: image processing -- techniques: interferometric -- planets and satellites: surfaces
\end{keywords}



\section{Introduction}

Io, one of Jupiter's largest moons, is the most geologically active body in the Solar System, with a heat flow $\sim$30 times more than that of Earth at present \citep{Matson_1981}. Its surface is dominated by hundreds of active volcanoes \citep[which produce roughly 55\% of Io’s heat flow;][]{Veeder_2015,davies2024}, driven by intense tidal forces generated by its Laplace (4:2:1)
orbital resonance with Europa and Ganymede \citep{Peale_1979}. This tidal heating causes Io's interior to flex and melt, fueling its persistent volcanic activity. The surface is covered with vast lava flows, calderas, and volcanic pits, with some lava fields stretching for hundreds of kilometers. Sulfur and sulfur dioxide frost blanket large areas, giving Io its distinctive yellow, red, and white hues \citep{Sagan_1979}. Frequent volcanic eruptions deposit new material across the surface, rapidly reshaping its landscape \citep{Davies_2007, Lopes_2007, de_Pater_2021}. Additionally, Io lacks significant impact craters, as its surface is constantly renewed by volcanic processes. The thin atmosphere, primarily composed of sulfur dioxide, further highlights Io's extreme and dynamic geological environment.

The volcanic activity of Io was discovered with the imaging of the moon by \textit{Voyager 1} and \textit{Voyager 2} \citep{Morabito_1979, Smith_1979}. Since then, a large collection of high-resolution observations of hot spots and plumes have been obtained by spacecrafts \citep[e.g., ][]{Rathbun_2014,davies2024}, by adaptive optics (AO) techniques on 8-10 m class ground-based telescopes \citep[e.g., ][]{marchis2002,depater2014Io, de_Kleer_2016b, de_Kleer_2019}, and interferometric radio observations \citep{de_Pater_2020,redwing2022,dekleer2024}. In this context, the James Webb Space Telescope (JWST) offers a unique opportunity to analyze potential changes on the surface or atmosphere of Io \citep[see e.g.,][]{de_Pater_2023}. Furthermore, JWST offers, for the first time, the opportunity to perform Fizeau interferometry with a space-based mission \citep{Artigau_2014}. 

\begin{figure}
    \includegraphics[width=0.98\linewidth]{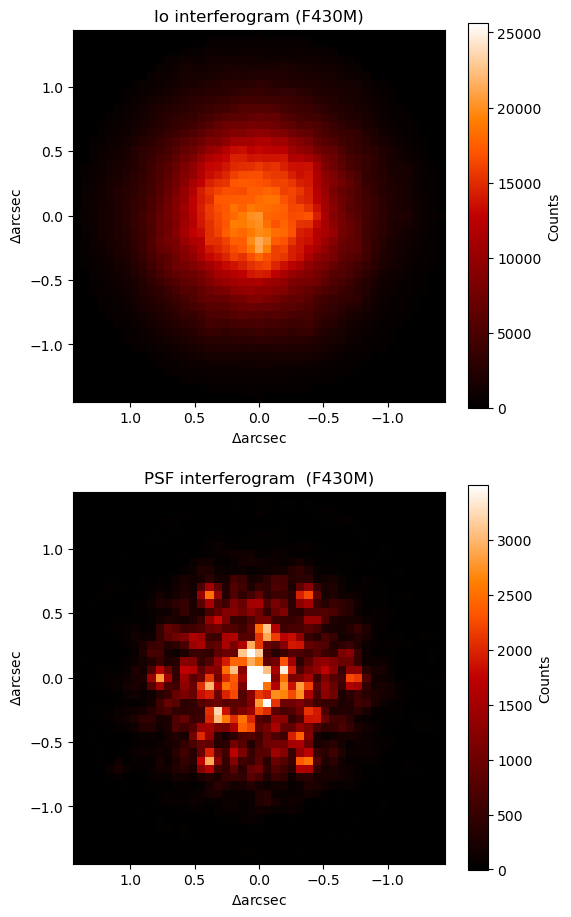}
    \caption{One of the integrations of Io (upper panel) and of the PSF (lower panel) after the data reduction process. Notice how the interference pattern of the PSF is crisp, while the interferogram of Io is blurred due to the extended nature of the target.}
    \label{fig:IO_interferograms}
\end{figure}

Aperture Masking Interferometry \citep[AMI][]{Tuthill_1999, Lacour_2004, Sanchez-Bermudez_2014} is a high-angular-resolution imaging technique that transforms a single telescope into an interferometric array. This method involves placing a non-redundant mask at the pupil plane of the telescope, which selectively transmits light through a sparse array of apertures. The resulting interference pattern, or interferogram, captures information about the spatial structure of the observed object at a maximum resolution of $\theta$ = $\lambda_0$/2B, where $\lambda_0$ is the central wavelength of observations and B the maximum separation between the holes of the non-redundant mask when projected on the primary mirror. By acting as a spatial filter AMI allows us selectively resolve specific spatial scales of interest in celestial objects that are otherwise blurred in conventional imaging techniques. During the last years, AMI has gained interest in the community. Several infrared cameras on ground-based 8-m class telescopes integrate this mode of observation. For example SPHERE and VISIR at the Very Large Telescope and \citep[VLT;][]{Cheetham_2016, Blakely_2022, Isbell_2022}, and the Large Binocular Telescope Interferometer \citep[LBT;][]{Sturmer_2012}. It has also been used successfully from space-based observations with the JWST \citep{Sallum_2024, Lau_2024, Blakely_2025}. However, the Io data analyzed in this work pushes the technical capabilities to the limit at observing a target as big as the interferometric field-of-view (FOV) of AMI. At the same time, it will allow us to observe with a resolution of 84 mas ($\sim$ 264 km at a distance of 4.363 astronomical units -au-) at mid-infrared wavelengths ($\lambda_0$ = 4.3 $\mu$m). This resolution is close to the measured sizes of several known volcanic structures on the surface of Io \citep[e.g. the Loki Patera is 200 km in diameter;][]{Rathbun_2006}, which allows us to identify individual sources.

The two main methods to analyze AMI data are: a) Fourier inversion of the interferogram, which will produce a series of "splodges" from which the interferometric observables (squared visibilities and closure phases) are computed. The main software that employs this method is \texttt{AMICAL} \citep{Soulain_2020}. b) Fitting the fringes of the interferogram directly to get their contrast and relative position of the phase reference and their corresponding observables. Some algorithms that use this method includes \texttt{ImplaneIA} \citep{Greenbaum_2015} and \texttt{SAMPip} \citep{Gallenne_2022, Sanchez-Bermudez_2022}. 

Another method to analyze AMI data consists of deconvolving the interferogram. This process involves reversing the effects of convolution introduced by the telescope's point spread function (PSF). Deconvolution algorithms iteratively refine the observed data to produce a clearer and more accurate representation of the object's true structure. Techniques such as the Lucy-Richardson \citep[L-R;][]{Lucy_1974} deconvolution and Maximum Entropy Method \citep[MEM;][]{Narayan_1986} are commonly used to recover high-fidelity images from the complex interferometric data. However, some of these methods have several limitations that affect the quality of the final restored image, particularly when the image contains extended structures. For example, L-R deconvolution favors the formation of point-like sources. Therefore, when there is extended emission, the algorithm tends to create artificial point-like objects instead of a continuous emission. 

The limitations on existing algorithms justify exploration of new techniques for image deconvolution for AMI observations.  Here we explore the use of artificial Neural Networks \citep[NN;][]{Bishop_1995, Egmont-Petersen_2002}. Neural networks are a class of machine learning models inspired by the structure and function of the human brain. They consist of interconnected layers of nodes, or neurons, where each connection represents a weight where knowledge is stored through a learning process. The power of neural networks lies in their ability to learn complex, non-linear mappings from inputs to outputs. During training, the network adjusts their internal variables, called weights, to minimize a loss function, which quantifies the difference between the predicted and actual outputs. This iterative optimization allows neural networks to capture intricate patterns in the data, making them highly effective for tasks such as image recognition \citep{Flamary_2017}, natural language processing \citep{Wang_2018}, and time series prediction \citep{Jamal_2020}.

Convolutional neural networks \citep[CNNs;][]{Lecun_1980} are well-suited for image processing due to their ability to learn hierarchical features from data. When applied to deconvolution, neural networks can directly learn the mapping from blurred images to their sharp counterparts by training on large datasets of paired images. This data-driven approach enables neural networks to implicitly model the deconvolution process, including bypassing the need for explicit PSF estimation.

In this work, we analyze neural network deconvolved images of Io at 4.3 $\mu$m, obtained with the AMI mode of the NIRISS camera aboard JWST. Sect. \ref{sec:observations} describes the observations. Sect. \ref{sec:analysis} presents different deconvolution techniques, the images obtained and their analyses. Our results are discussed in  Sect. \ref{sec:discussion} and the final Section summarizes our conclusions. 

\section{Observations }
\label{sec:observations}

\begin{table*}
\caption{Summary of the JWST/NIRISS AMI observations of Io and the calibrator HD\,2236}
\begin{threeparttable}[b]
\label{tab:observation}
\begin{tabular}{|| c c c c c c  c c||}
 \hline
Starting time$^{1}$ & Object & Obs. Sub-Long. & Obs. Sub-Lat. & Exp. Time [sec]  & NGROUPS & NINTS & Dither Pattern\\
 \hline
 16:32:23 UT  & Io & 100.682 & 2.659 & 349.072 &  45 & 100 & Stare / sub-px. dither \\
 16:40:56 UT & Io & 101.891 & 2.659 & 349.072 & 45 & 100 & Stare / sub-px. dither \\  
 16:49:29 UT & Io & 103.099 & 2.659 & 349.072 & 45 & 100 & Stare / sub-px. dither \\
16:58:02 UT & Io & 104.308 & 2.659 & 349.072 & 45 & 100 & Stare / sub-px. dither \\
 17:06:35 UT & Io & 105.374 & 2.631 & 349.072 & 45 & 100 & Stare / sub-px. dither \\
 17:53:37 UT & HD\,2236 & - & - & 8.010 & 12 & 8 & 4-point dither\\
 17:54:57 UT & HD\,2236  & - & - & 8.010  & 12 & 8 & 4-point dither\\
 \hline
\end{tabular}
\begin{tablenotes}
       \item [1] All observations were taken on August 1st., 2022
     \end{tablenotes}
 \end{threeparttable}
\end{table*}

As part of the JWST Early Release Science program \#01373 \citep{depater2022JWSTDPS} we observed the Jovian satellite Io with the JWST NIRISS instrument \citep{Doyon_2023} in its Aperture Masking Interferometric mode \citep[][]{2010SPIE.7731E..3WS,Sivaramakrishnan_2023}. Observations of Io and PSF calibrator star (HD\,2236) were taken using the F430M ($\lambda_{0}$ = 4.285 $\mu$m, $\Delta \lambda$ = 4.182 - 4.395 $\mu$m) filter on August 1st, 2022. The observations were taken using the standard AMI parameters with the “NISRAPID” readout pattern and the SUB80 (80 × 80 pixel) subarray, where the NIRISS detector plate scale is 65.7 mas pixel$^{-1}$. Five science exposures were obtained for Io. Each exposure (or data cube) corresponds to a different sub-pixel dither position. The science exposures consist of 45 groups per integration and with 100 integrations per exposure. Four calibrator exposures (using a 4-point dither pattern) were observed with 12 groups per integration and 8 integrations per exposure. The effective exposure time for each of the five science exposures is 338.48 seconds and the effective exposure time for each of the 4 calibrator exposures is 7.23 seconds.


The data were processed using the standard data reduction \texttt{Python jwst} pipeline\footnote{\url{https://jwst-pipeline.readthedocs.io/en/latest/getting_started/quickstart.html}} and \texttt{CALAMITY}\footnote{\url{https://github.com/rcooper295/CALAMITY}} to obtain level 2 data products. The latter is a small \texttt{Python} wrapper, with a simple but functional command line interface, which allows us to determine the level of data reduction products required. These algorithms produced interferograms corrected by cosmetics and integrated across 45 and 12 groups for the science and calibrator, respectively. While the NIRISS-AMI threshold of the detector linear response is of 26000 counts, in order to avoid charge migration problems at the brightest pixels, we used a more stringent threshold of 16227 counts (Sivaramakrishnan priv. comm.). The bad pixel correction was performed using \texttt{Astro-Fix}\footnote{\url{https://github.com/HengyueZ/astrofix}} based on the Gaussian process regression described by \citet{Zhang_2021}. The interferograms were centered and cropped to grids of 45 $\times$ 45 pixels. A summary of the data used in this work is provided in Table \ref{tab:observation}. Figure \ref{fig:IO_interferograms} displays the interferograms for both the PSF and Io for one of the integrations in the data. 


\section{Analysis and Results }
\label{sec:analysis}

\begin{figure}
\centering
    \includegraphics[width=0.8\linewidth]{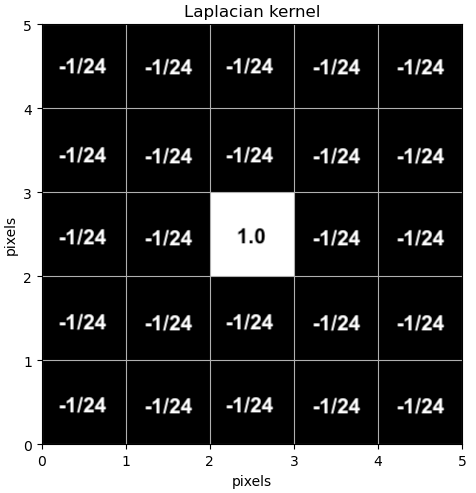}
    \includegraphics[width=0.95\linewidth]{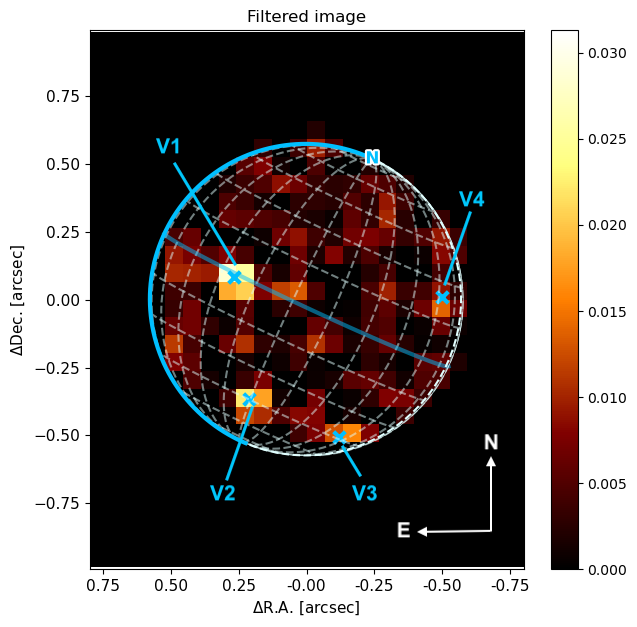}
    \caption{The upper panel shows a graphical representation of the kernel used for Laplacian-like filtering. The value of the 5 $\times$ 5 pixel kernel are labeled in the panel. The lower panel shows the filtered median image of the first data cube in the observations. The flux centroid of the four brightest features as defined in Table\,\ref{tab:volcanoes_fluxes} are labeled with blue crosses. The orientation of the North pole is labeled, the equator and the meridian on the morning limb are shown with blue solid lines and meridians are plotted with white dashed-lines every 15 deg. The image shown is filtered using a circular mask with a diameter of 18 pixels. The emission observed is normalized to unity and the scale is shown in the panel.}
    \label{fig:filtered_im}
\end{figure}

\subsection{Laplacian-like filtering for deconvolution}

As shown in Figure \ref{fig:IO_interferograms}, the contrast of the individual fringe patterns produced by each one of the baselines in the non-redundant mask (NRM) is quite low; this causes Io's interferograms to be extremely blurred. This makes it difficult to use standard algorithms based on fringe fitting \citep{Sanchez-Bermudez_2022, Lau_2024} or numerical Fourier transformation \citep{Soulain_2023} to extract the interferometric observables (squared visibilities and closure phases) to reconstruct the correct brightness distribution of the object. Therefore, we decided to explore deconvolution methods to extract Io's projected morphology in the plane of the sky. This requires us to remove (or deconvolve) the NRM's point-spread-function from the science images. As a first test, we used a 'center-surround' convolutional Laplacian-like filter to isolate sharp features in the AMI interferograms. This method was inspired by retinal processing found in mammalian vision, for example, in cat retina ganglion cells \citep{Hubel_Wiesel_1959}. 

Prior to JWST's launch, \citet{2016LPI....47.3005T} investigated various center-surround filter geometries to demonstrate feasibility of the observational strategy used to collect AMI data, and to develop an easy and fast method to reveal compact structures on the surface of Io. One of the best method for this turned out to be the Laplacian-like kernel. The Laplacian operator measures the second spatial derivative of an image. For point-like sources, which often appear as small, bright (or dark) regions contrasting with their surroundings, the intensity changes rapidly. Thus a Laplacian-like filter will show a zero-crossing or a strong peak (either positive or negative depending on the kernel's definition) at these points, highlighting them \citep{Woods_2012}. Another good kernel for highlighting point-like objects is the Difference of Gaussians -DoG- \citep[see e.g.,][]{Marr_1980}.

Our objective of using this Laplacian-like simple linear filter is to use the identified structures on Io's disk as proxy for more advanced image reconstruction techniques, where kernels are learned automatically. The used finite support filter possesses a single pixel positive core surrounded by negative wings. The extent of these negative wings is set to the scale of the background structure, with the positive peak width setting the scale of interesting features. Our implementation consisted of a unity response positive central pixel in a 5 x 5 pixel kernel, with all the other 24 pixels set to -1/24 (see the upper panel in Fig.\,\ref{fig:filtered_im}). The kernel shifts over the input image, convolving each overlapping image section with the kernel matrix. This ensures a null response to uniform illumination and an insensitivity to intensity gradients in the image.  For comparison, relevant scales in the AMI image are the 16-pixel diameter of Io's disk, the approximately 1 pixel (Michelson criterion) resolution of JWST AMI at the observing wavelength, and the 16-pixel primary beam diameter (the interferometric field of view). The lower panel in Fig.\,\ref{fig:filtered_im} displays the median filtered image obtained with the first science exposure. The moon's structure over the disk shows four bright spots V1-V4 with a signal above 40\% of the image's peak, below this value background fluctuations are of similar level. The brightest spot, V1, is located close to the equator line of the projected moon's disk at the time of the observations. It is important to highlight that the Laplacian-like kernel produces negative pixel values in the image. Thus, the flux calibration and the background flux estimation (particularly for faint pixel values, is limited with this technique.

\begin{figure}
    \includegraphics[width=\linewidth]{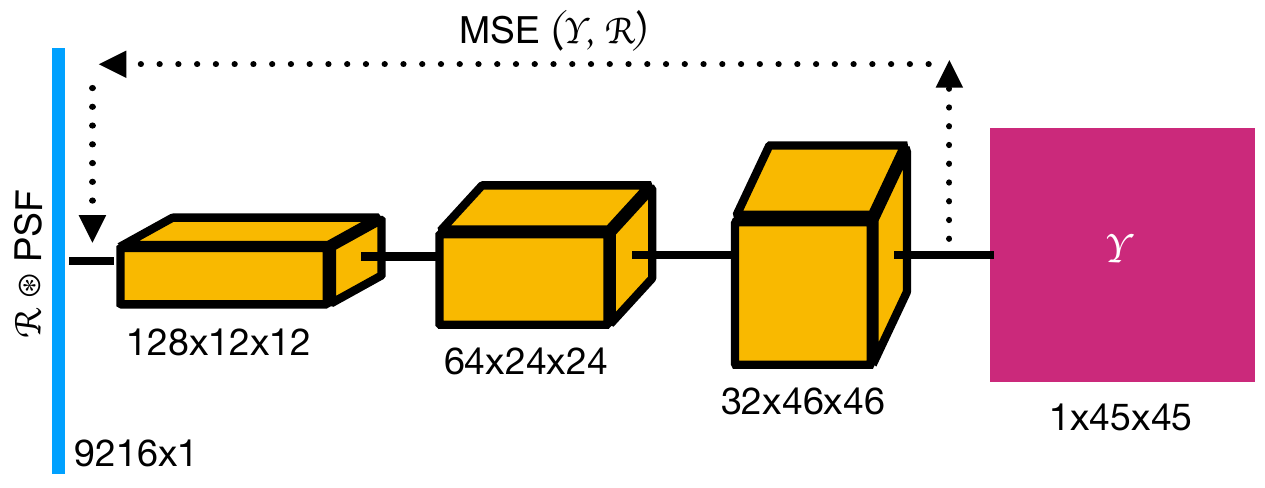}
    \caption{Schematics of the CNN architecture. (i) The yellow cubes correspond to the hidden convolutional layers; (ii) the blue vector represents the initial dense layer, which receives the input convolved model $R \circledast \mathrm{PSF} $ and; (iii) the magenta rectangle represents the output image. The recovered image, $Y$, is compared with the unconvolved model, $R$. The dimensions of each layer and of the output image are labeled in the diagram.}
    \label{fig:diagram1}
\end{figure}

\begin{figure*}
    \includegraphics[width=\linewidth]{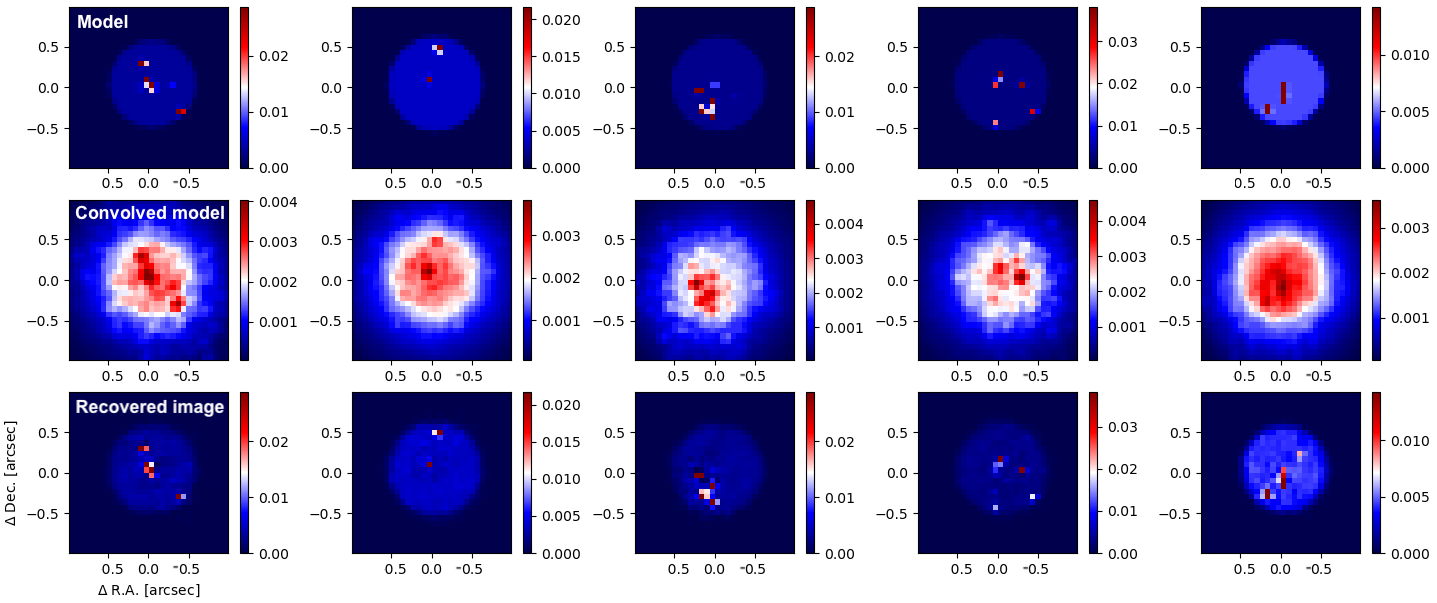}
    \caption{Examples of the injection-recovery experiment for the CNN. The first row displays five models from the validation library. The second row displays their corresponding convolved (input) images. The third row displays the recovered images from the CNN. All the color scales are normalized and the first and third rows are shown with the same range of values. The FOV of all panels is the same. }
    \label{fig:CNN_ims2}
\end{figure*}

\subsection{Image deconvolution with neural networks}

\subsubsection{Supervised convolutional neural network}
\label{sec:cnn}
To improve on the Lapacian-like filtering, we designed a specific supervised NN which is based on an architecture conformed by convolutional layers to deconvolve Io's data. It has been shown in the past that this kind of networks are good at recovering images, even from sparse measurements in the sensor domain. For example, in \citet{Sanchez-Bermudez_2022} it was demonstrated that a CNN performs well at identify structures from interferometric observables to recover an image \citep[see also ][]{Sanchez-Bermudez_2024}. In our case, given that it is extremely difficult to extract squared visibilities and closure phases from Io's data (given the extended disk emission of the moon), we followed a deconvolution approach with a CNN that takes advantage of the well-defined PSF constrained by the NRM geometry. The network was implemented using the \texttt{Python pytorch} library. The design of the CNN\footnote{In Apendix \ref{sec:glossary}, we included a glossary of technical terms used in the architectures of the NNs in Sects. \ref{sec:cnn} and \ref{sec:dip}. } considers (i) one dense input layer connected to (ii) three hidden convolutional layers and (iii) one convolutional output layer (see Fig.\,\ref{fig:diagram1}). A convolutional layer is commonly used for image processing with neural networks. The input of these layers consists of a multidimensional array that represents an image or feature map. One of the key components of these layers are learnable filters (or kernels). These filters are typically smaller in spatial dimensions (height and width) compared to the input. Each filter registers the input image, and computes dot products between the filter's weights and a local region of the input. This is, essentially, the same process as the one performed with the Laplacian-like filtering. Although, here, it is generalized to the case of learnable filters with different sizes. 

As the filter slides over the input, it produces a 2D activation map or feature map. This map highlights regions of the input that are activated by the filter, capturing important features such as edges, textures, or other patterns. After the filter registers the whole pixel grid of the input, an activation function is typically applied. This allows non-linearity to be introduced into the model and to learn more complex patterns. The output of the convolutional layer is a set of feature maps, one for each filter. In our architecture, all the hidden convolutional layers have filters of 4 $\times$ 4 pixels and the number of filters are 128, 64 and 32 for the first, second, and third hidden layer, respectively. A \texttt{ReLU} activation function is used for all the layers, except for the output one. For this last layer, we created a trainable activation function, called \texttt{AReLU}. This activation function is defined as follows:

\begin{equation}
    \textbf{Y} = \mathrm{maximum} \left[A \textbf{X}, \textbf{X} \right]\,,
\end{equation}

where $\textbf{X}$ is the matrix output of the third hidden layer and $A$ is a trainable weight. The weight is initialized with a random number from a Normal distribution.  The CNN receives a convolved image or model as input (reformatted as a 1D column vector) and outputs a 2D generated image. The CNN needs to be trained using a library of models provided by the user to learn the most representative features in the images. In this case, we create 30000 different model images. Each model considers a uniformly illuminated disk with bright pixels dispersed across it. The size of the disk is varied between 16 and 18 pixels. While the number (and positions) of bright pixels on the surface of the disk varies randomly between 1 to 10. The flux of the disk varies between 80\% and 95\% of the total flux in the image, and the rest is distributed among the corresponding bright pixel elements in the surface. To create the corresponding pair of convolved images, the models were convolved randomly with one of the calibrator PSFs. All the model images were cropped to pixel grids of 45 $\times$ 45 pixels. The CNN was trained for 2000 epochs, after which we observe convergence. For each epoch, for a given input model the generated image by the CNN is compared with the corresponding convolved model using the Mean Squared Error (MSE) of the pixel differences as objective function\footnote{The \texttt{Python} code of our CNN can be consulted at the following \texttt{Github} link: \url{https://github.com/cosmosz5/Io_CNN.git}}.

\begin{figure}
    \includegraphics[width=9 cm]{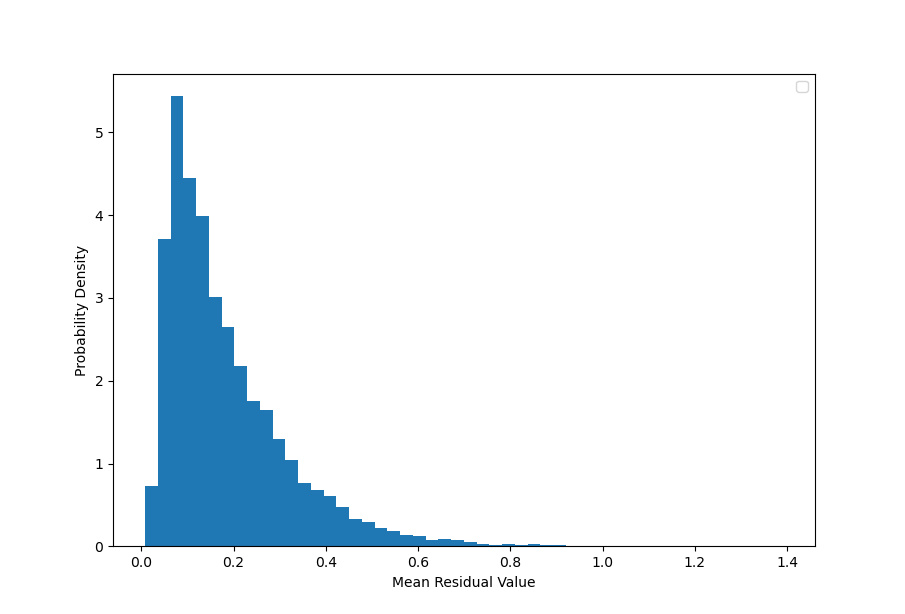}
    \caption{Probability density distribution of the mean residual values of the images generated from the validation library of models for the CNN.}
    \label{fig:residuals}
\end{figure}

To verify the results after trainning the CNN, we validated it with an injection-recovery experiment. For this, we created additional 10,000 models from which images were recovered. Fig. \ref{fig:CNN_ims2} displays, in the first and second rows, 5 pairs of deconvolved-convolved models from this validation library. The third row in Fig. \ref{fig:CNN_ims2} displays the output images from the CNN. As shown, the recovered images are consistent with the structure of the respective deconvolved models. This demonstrates that the CNN was properly trained and that the injected bright sources could be retrieved. To test systematically the validity of the recovered images, we computed the average pixel residual per recovered image of the validation library. Fig. \ref{fig:residuals} shows a histogram of the mean residuals per pixel for the validation models. The peak of the average residual is around 10\% of the pixel value.      

\begin{figure*}
    \includegraphics[width=14 cm]{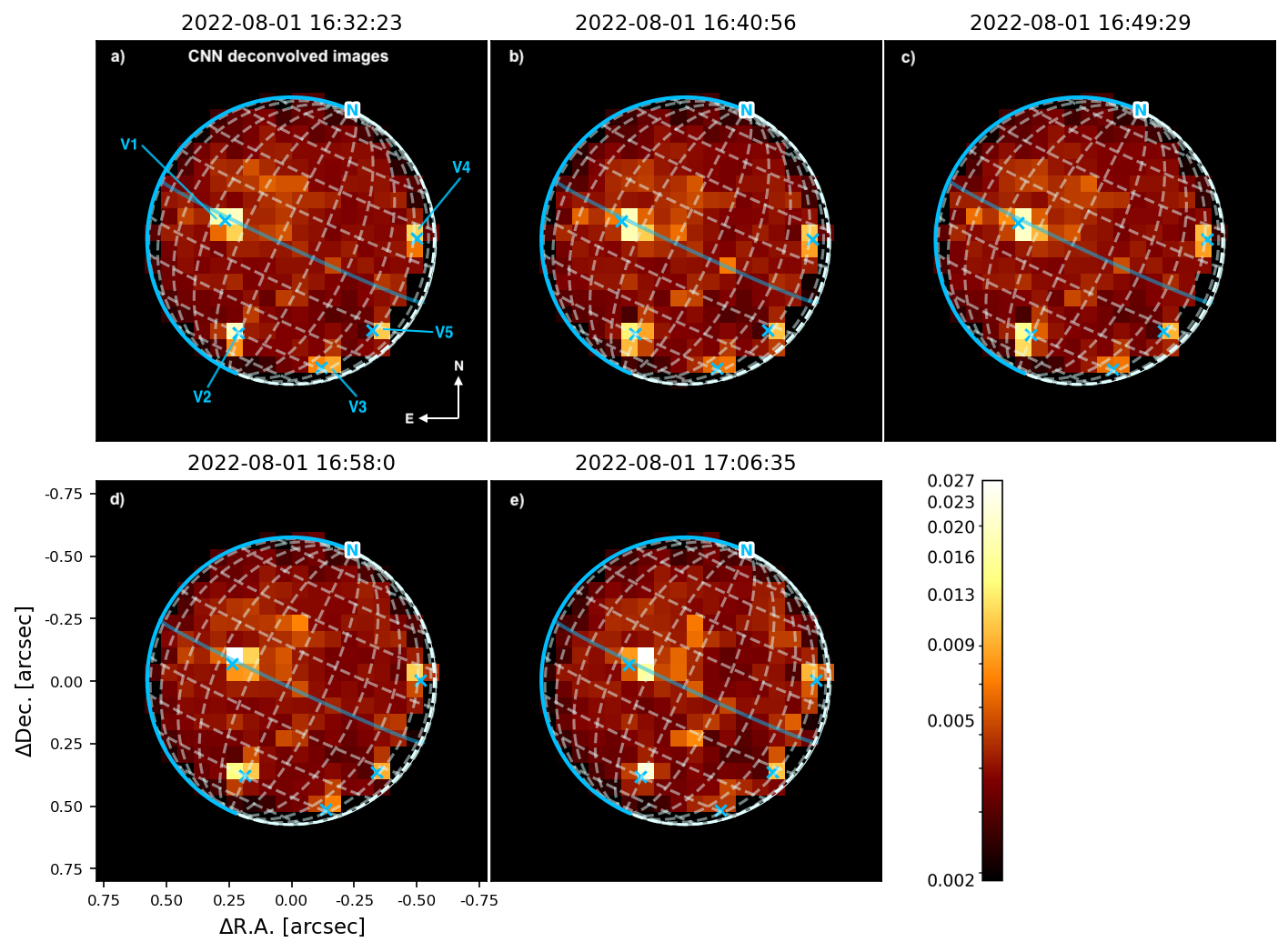}
    \caption{Median Io images per data cube of the JWST NIRISS-AMI observations, with superposed projected moon's disk. The orientation of the North pole is labeled, the equator and meridian on the morning limb are shown with blue solid lines and meridians are plotted with white dashed-lines every 15 deg. The grid of the projection is fixed to the 2022-08-01 16:32:23 ephemeris, so that moon's changes are observed relative to this date. The position of five volcanoes in the moon's surface are labeled with blue crosses and labels are displayed in panel a) as defined in Table\,\ref{tab:volcanoes_fluxes}. The image shown is filtered using a circular mask with a diameter of 18 pixels. The total emission observed is normalized to unity. The emission's scale is shown in the image and it is common to all the panels.}
    \label{fig:ims_CNN}
\end{figure*}

Finally, we used the AMI data to obtain the deconvolved images for each one of the integrations in the science exposures. Figure \ref{fig:ims_CNN} displays the median deconvolved images obtained with the CNN of each data cube. The images were derotated by the corresponding parallactic angle described in the data's header keyword \texttt{ROLL\_REF}. Notice that there are at least five bright spots (V1-V5) across the moon's surface, all of them with a signal above 40\% of the flux's peak. Below this value the background fluctuations are of the same scale in the deconvolved images. Four of them (V1-V4) are consistent with the ones observed in the Laplacian-like filtered images. 

\subsubsection{Unsupervised Deep Image Prior neural network}
\label{sec:dip}

\begin{figure}
    \includegraphics[width=\columnwidth]{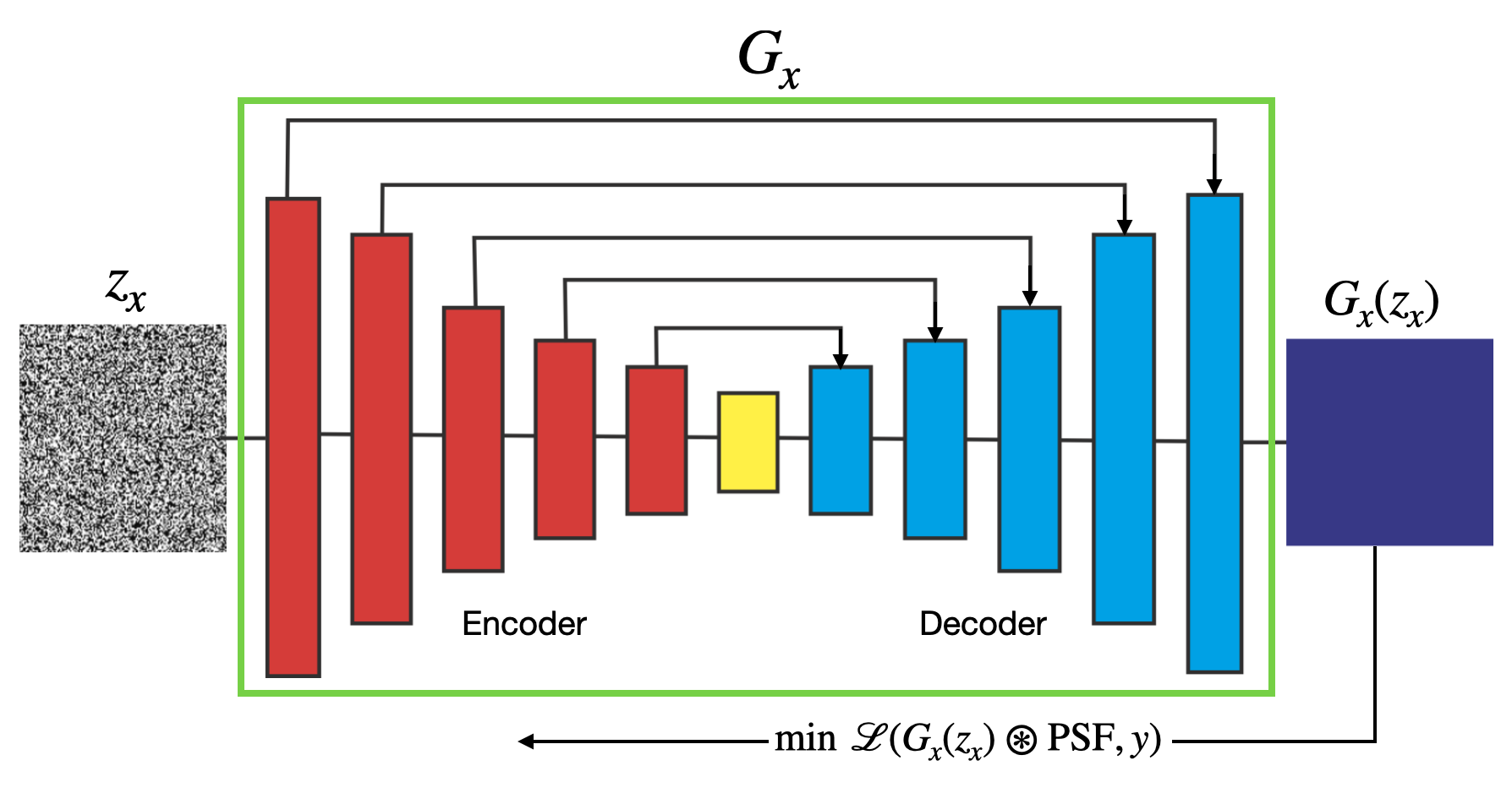}
    \caption{Schematics of the DIP network architecture. The red rectangles correspond to the layers of the Encoder; the yellow rectangle represents the latent space and; the blue rectangles represent the Decoder part of the networks. The Recovered image, G$_x$(z$_x$), is convolved with the PSF before being compared with the ground truth image $y$.}
    \label{fig:VAE_architecture}
\end{figure}

Despite the fact that the CNN is able to obtain reasonable deconvolved images, we realized that the output of this network is highly limited by the models used during training. This constrains the performance of the network when structures that are not initially considered in the models are present in the data. For example, the CNN only considered point-like sources and not extended emission. In order to improve our results, we tested a generative neural network with the general architecture of an autoencoder. This is a type of unsupervised forward model that represents high-dimensional data in a lower-dimensional latent space from which a model of the original image is recovered. In this case, we follow the Deep Image Prior (DIP) algorithm described by \citet{Ren_2019} and \citet{Ulyanov_2020}, in which the autoencoder serves as a denoiser, in which the input is Gaussian noise and the output the deconvolved image. A schematic of the network is shown in Fig. \ref{fig:VAE_architecture}. The main components of the network are also convolutional layers. However, since one of the main limitations of the convolutional layers is that they struggle to learn correlations between a series of pixels that are not adjacent,the network incorporates skip connections across the layers. 

\begin{figure*}
    \includegraphics[width=14 cm]{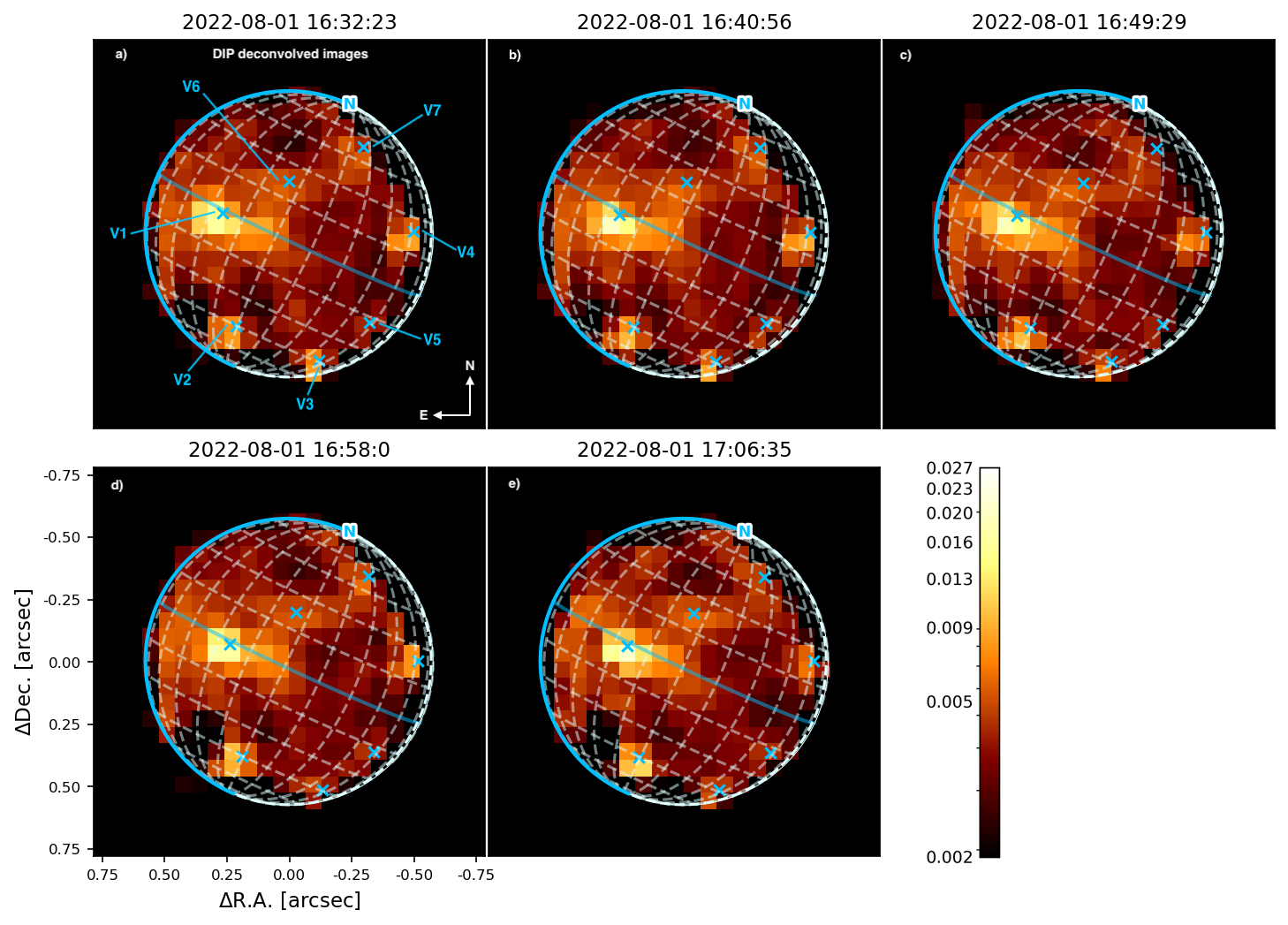}
    \caption{Median Io images per data cube of the JWST NIRISS-AMI observations obtained with the DIP network, with superposed the projected moon's disk. The orientation of the North pole is labeled, the equator and the meridian on the morning limb are shown with blue solid lines and meridians are plotted with white dashed-lines every 15 deg. The grid of the projection is fixed to the 2022-08-01 16:32:23 ephemeris, so that moon's changes are observed relative to this date. The position of seven volcanoes on the moon's surface are labeled with blue crosses and labels are displayed in panel a) as defined in Table\,\ref{tab:volcanoes_fluxes}. The image shown is filtered using a circular mask with a diameter of 18 pixels. The emission observed is normalized to the unity. The emission's scale is shown in the image and it is common to all the panels.}
    \label{fig:Io_ims}
\end{figure*}

For the encoder part of the network, each compression block before a skip connection consists of two sequences of (i) a convolutional layer with a stride of two, 128 feature maps, and a filter size of 3 $\times$ 3 pixels; (ii) a \texttt{Batch Normalization} layer; (iii) a \texttt{Reflection Padding 2D} layer to take into account the information in the borders of the original image and; (iv) a \texttt{LeakyReLU} activation layer, which favors positive linear models. The skip connection block uses a sequence of (i) a convolutional layer with a filter size of 1 $\times$ 1 pixels, 16 feature maps and a stride of one; (ii) a \texttt{Batch Normalization} layer and; (iii) a \texttt{LeakyReLU} layer. Finally, for the decoder part of the network, each decompression block before reconnecting a skip connection consisted of a \texttt{Batch Normalization} layer, followed by two sequences of (i) a convolutional layer with a stride of one, 128 feature maps, and a filter size of 3 $\times$ 3 pixels; (ii) a \texttt{Batch Normalization} layer and; (iii) a \texttt{LeakyReLU} activation layer. A final \texttt{Upsample} layer with a scale factor of two is added. Both the encoder and decoder consisted of 5 concatenated compression or decompression blocks, respectively.

We used, as base model, the DIP network implemented by \citet{Ren_2019} coded in \texttt{Pytorch}\footnote{The DIP base code can be accessed at the following link: \url{https://github.com/csdwren/SelfDeblur}} and we modify it to the specific purpose of Io. The training was executed in a Mac Studio with an M2 Ultra processor and 64 Gb of shared memory. Each integration of the science exposures was deconvolved individually. For this, an Adam optimizer was used with an initial learning rate of $\mu$ = 0.001 and and adaptive coefficient $\gamma$ = 0.99 with milestones at 5,000, 10,000, 20,000, and 40,000 iterations. The DIP-NN uses a noise-based regularization, which adds normal noise with zero mean and a standard deviation $\sigma_p$ to the input signal $z_x$. In this case, we use $\sigma_p$ = 1/30. The total number of iterations per integration was of 40000. The objective function used is the complement of the structural similarity index measure (SSIM): 1 - SSIM, between the ground truth, \textbf{$y$}, and the restored image, $R_x$, which is defined as: 

\begin{equation}
    R_x = G_x (z_x) \circledast \mathrm{PSF} \,,
    \label{eq:conv}
\end{equation}

where $G_x (z_x)$ is the deconvolved image produced by the DIP network and the PSF is a random integration obtained from the calibrator datasets. SSIM is a metric that asses the quality of an image using the perceptual characteristics of the human visual system. This metric evaluates three main features from the image: the luminance, contrast and structure. The measure between two images \( y \) (the reference image) and \( R_x \) (the sample image) of common size \( N \times N \) is:

\[
\text{SSIM}(R_x, y) = \frac{(2 \mu_{R_x} \mu_y + C_1)(2 \sigma_{R_x,y} + C_2)}{(\mu_{R_x}^2 + \mu_y^2 + C_1)(\sigma_{R_x}^2 + \sigma_y^2 + C_2)}\,,
\]

where $\mu_y$ and $\mu_{R_x}$ are the means of the reference and sample images, respectively. $\sigma_{R_x}^2$ and $\sigma_{y}^2$ are the variance of the pixel values in the images, while $\sigma_{R_x,y}$ is the covariance between the two images. The coefficients C$_1$ and C$_2$ serve to stabilize the division with weak denominator, they are defined as C$_1$  = (k$_1$ L)$^2$ and C$_2$  = (k$_2$ L)$^2$. L is the dynamic range of the pixel-values while k$_1$ = 0.01 and k$_2$ = 0.03. 

 The final image per science exposure was obtained by calculating the median over the integrations in the corresponding exposure and derotated (see Figure \ref{fig:Io_ims}). At least four bright spots are identified on the surface of Io (with a signal above 40\% of the emission's peak), with three fainter ones (with a signal above 25\% of the emission's peak). Below those values, background fluctuations in the DIP images are of similar scale. They are identified as V1-V7 and are labeled in the first panel of Fig.\ref{fig:Io_ims}. Notice that the moon's disk was also restored without any specific prior as in the case of the CNN. 

\subsection{Keck Observations Prior to and Following the JWST Observations}

\begin{figure*}
    \includegraphics[width=18 cm]{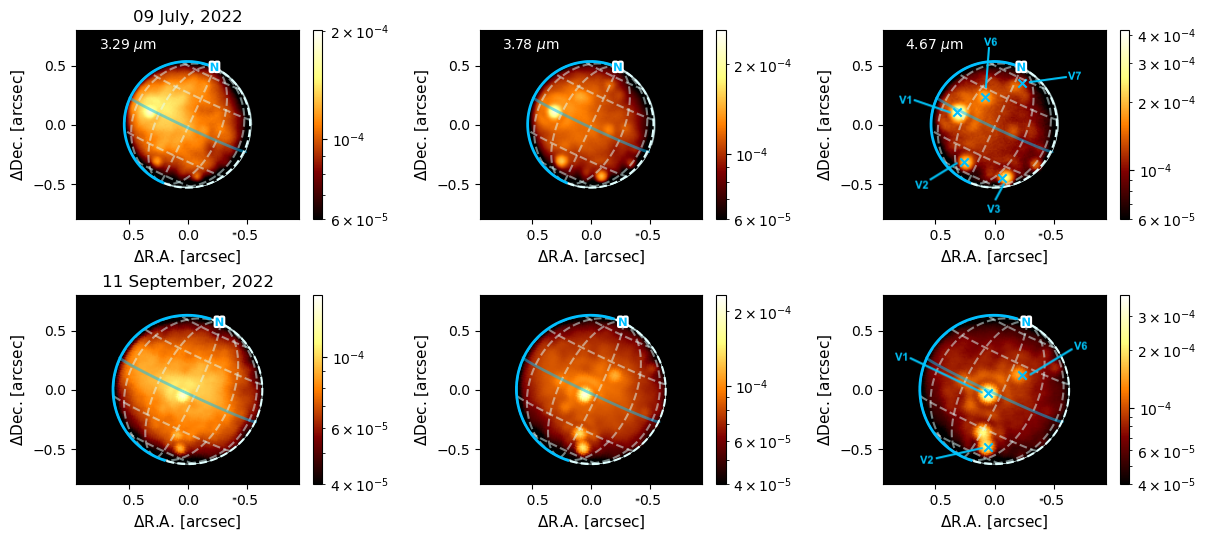}
    \caption{Images of Io taken with the Keck telescope on 9 July and 11 Sep. 2022. The columns show the different filters at which the images were taken (see label on the panels) and the rows correspond to the two different epochs. The moon's disk projected at the time of the observations are over-plotted. The orientation is the same as for the JWST AMI images. The emission of each panel is normalized to the unity and emission scale is shown in the different color-bars of each panel.  }
    \label{fig:keckIo}
\end{figure*}

To better understand the structure of Io and the bright spots observed in the deconvolved AMI data, we complemented the JWST data with ground-based Adaptive Optics (AO) data. Io was observed with the Near-Infrared camera (NIRC2) coupled
to the AO system on the Keck II telescope
\citep{wizinowich2000} at a similar viewing geometry roughly a month before and after the JWST observations, on 2022 July 9 and September 11.
NIRC2 is a 1024 $\times$ 1024 Aladdin-3 InSb array, which was used with the NARROW camera which has a plate scale of
9.971 $\pm$ 0.004 mas per pixel \citep{service2016}. Io itself was used for wavefront sensing. 
All images were processed using standard near-infrared data reduction
techniques \citep[flat-fielded, sky-subtracted, with bad pixels replaced
by the median of surrounding pixels; ][]{roe2002}. The geometric distortion
in the Keck images was corrected using the solution provided
by \citet{service2016}. Three images were taken in the two broadband filters, Lp (3.776 $\mu$m; $\theta$ = 97 mas) and Ms (4.67 $\mu$m; $\theta$ = 120 mas), as well as the narrow-band filter PAH (3.29 $\mu$m; $\theta$ = 85 mas). Each image was integrated over 18--20 seconds.
The individual images were aligned and co-added to increase
the signal-to-noise. Photometric calibration was bootstrapped from data taken on June 21, 2023, at a similar viewing geometry, when 
the standard star HD1160 was observed \citep{leggett2003}.
The observations taken on 2020 July 9 were taken as part of our Twilight Zone Program \citep{depatertwi}. The images are shown in Figure \ref{fig:keckIo}.

The viewing geometry for the 2022 July 9 Keck images was very similar to that of the JKWST-AMI images. Hence we see the brightest spot close to the location of the brightest spot in the JWST-AMI deconvolved images (obtained with the three methods: the Laplacian-like filter, CNN and DIP-NN). Two of the bright spots at the bottom of the moon's disk and very close to the limb are also present in both the Keck July images and on the JWST ones. The extended emission observed to the right of the brightest spot in the Keck images is more diffuse than the one observed in the JWST-AMI DIP deconvolved images.

\section{Discussion}
 \label{sec:discussion}

 \subsection{Comparison between the deconvolution methods}

 The different deconvolution methods presented in this work show distinct advantages and disadvantages for analyzing AMI data. The Laplacian-like filtering method offers a quick and easy way to isolate sharp, point-like features in the images. However, a significant drawback of this method is its production of negative pixel values, which in turn limits the accuracy of flux calibration and background flux estimation. The CNN is adept at recovering images, even from sparse data, by learning complex, non-linear mappings from the input data. Our CNN architecture was specifically designed to handle the difficulties of extracting interferometric observables (from standard fringe fitting and Fourier Transform interferometric analysis) from Io's data due to its extended disk emission. It offers a more robust and sophisticated method for image reconstruction, capable of handling the complexities of the dataset at the cost of more computational resources and the necessity of a large model collection for training. Finally, the DIP network is an unsupervised method that does not require a pre-existing training dataset. Instead, it uses the structure of the network itself as a prior to guide the image reconstruction process. This makes the DIP particularly useful when a representative training dataset is not available.
 
While both CNN and DIP can produce high-quality deconvolved images from AMI data, the key difference lies in their learning paradigm: the CNN learns from a large dataset of examples, while the DIP leverages the network's structure as an implicit prior to reconstruct the image from a single convolved (blurry) input. This distinction makes the CNN a powerful tool for specific, well-defined problems with ample training data (e.g., when radiative transfer or geometrical models are available), whereas the DIP offers a more flexible, data-independent solution for image reconstruction. In the following sections, we conduct a detailed comparison of the Keck images and the DIP ones, we also characterize the compact and extended structures observed in the DIP images.

\subsection{Bright spots and volcanic activity}

\begin{figure}
    \includegraphics[width=8 cm]{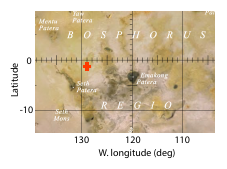}
    \caption{The eruption site, V1 (the red cross), indicated on a USGS map 
(\url{https://astrogeology.usgs.gov/maps/io-voyager-galileo-global-mosaics}).}
    \label{fig:USGS}    
\end{figure}

The bright spots in the JWST-AMI and Keck images are commonly referred to as ``hot spots'', which are areas of hot lava caused by the intense volcanic activity of Io. We relate the seven hot spots observed in the DIP images (Fig. \ref{fig:Io_ims}) with volcanoes on the surface of Io. The brightest eruption, V1, in these images occurred just NE of Seth Patera. The position of V1 is indicated on top of a United States Geological Service (USGS) map on Figure \ref{fig:USGS}. The position of V1 (129.2 $\pm$ 0.5$^\circ$ W. longitude, 2.0 $\pm$ 0.5$^\circ$ S. Latitude) was derived from the Keck image taken on 11 September 2022, where the hot spot was relatively close to the center of the disk, and hence the positional accuracy was highest. Navigation of the Keck image was easier than  of the original AMI images because the extended structure and the limb is well-defined in the Keck images. The radiant flux at that time was 77$\pm$8 GW/$\mu$m in Lp band and 130$\pm$15 GW/$\mu$m in the Ms band, as derived from the Keck images. 
This eruption was already ongoing on 2022 May 17, and continued through at least October 2023, as seen from Keck twilight images \citep{depatertwi} and {\it Juno}/JIRAM data \citep{Perry_2025}.

The position of V1 was then used to navigate the JWST-AMI images. After the naviagion an alignment of the DIP images, the final V1 position reported in Table \ref{tab:volcanoes_fluxes} is equal to the one derived with the Keck images within
the errorbars. To determine the position of the emission peaks of the other volcanoes, we projected the DIP images into a Lon.-Lat. map with a pixel scale of 1$^{\circ}$ (Fig.\,\ref{fig:projection_io} in the Appendix displays an example of this projection). From these projected maps we obtained the longitude and latitude of each of the hot spots, as summarized in Table \ref{tab:volcanoes_fluxes}. The reported errorbars are the standard deviation of the values averaged over the five science exposures. The derived position of V1 from the DIP images is equal to the one derived with the Keck images within the errorbars. Based upon these positions, we suggest that V2 most likely corresponds to P197, V3 to Masubi, V4 to N. Lei-Zi Fluctus, V5 to Kanehekili Fluctus, V6 to Amirani and V7 to Tonatiuh. We note, though, that the derived longitude of Kanehekili Fluctus is 10-15$^\circ$ off; however that region was extremely active in 2022--2023 \citep{de_Pater_2023}, and in the AMI data we see it essentially on the limb of the satellite. The AMI data show an eastward extension of the bright eruption near Seth Patera, which perhaps involves emission from Emakong Patera. Although this was not evident in the Keck data taken roughly one month before and after the JWST data, volcanic activity is highly variable over time. The hot spots V2 (P197), V3 (Masubi), V6 (Amirani) and V7 (Tonatiuh) are also visible on the Keck image taken on July 2022. On September 11, 2022 V1, V2 and V6 are also observed together with one bright unknown source close to the V2 position (see Fig.\,\ref{fig:keckIo}).

In order to better characterize the emission of the volcanoes in the DIP images, we calibrated the absolute flux in the reconstructed images. For this, we performed the following method: 

\begin{itemize}
    \item We obtained the total number of counts or Analog-Digital Units (ADUs) enclosed in the core of the interferograms in both the science and the calibrator. For this, we first apodized the interferograms using a super-Gaussian window (with an exponent of 4), considering a FHWM of $\lambda/\mathrm{D}$, where D is the size of the holes of the NRM. We obtained the Fourier Transform of the interferograms and measured the value of the DC component at zero spacing (u=0,v=0). It is expected that for a large object such as Io (which has a size almost of the interferometric FOV), some flux is spread outside the interferogram's core. Thus, to characterize the amount of flux of an extended source dispersed over the halo of the PSF, and to evaluate the effect of using a super-Gaussian window, we used the 30000 simulations used for the CNN. For these we applied the apodizing method described above and we compared the total extracted flux from the simulations with the one obtained from a theoretical PSF. We concluded that on average we are able to recover $\alpha$ = 92 $\pm$ 1 \% of the total flux for sources with the similar size of Io.    
    \item We convert the Calibrator's magnitude (M$_{\mathrm{HD\,2236}}$ = 6.20 $\pm$ 0.03 mag) into flux density (Jy) considering the JWST-NIRISS filter profile (F$_{\mathrm{Cal}}$ = 0.65 $\pm$ 0.02 Jy). With the total ADUs of the calibrator and science, we used the following formula to compute the flux density of the target:

    \begin{equation}\label{eq:fluxcal}
\begin{split}
F_{\rm{Io}}\,[\rm{Jy}] = \frac{F_{\rm{Cal}}\,[\rm{Jy}]}{F_{\rm{Cal}}(u=0,v=0)\,[\rm{ADU}]} 
\times \frac{F_{\rm{Io}}(u=0,v=0)\,[\rm{ADU}]}{\alpha}\,.
\end{split}
\end{equation}

The measured $F_{\rm{Io}}(u=0,v=0)\,[\rm{ADU}]$ was corrected by the $\alpha$ factor determined with the simulations in the previous point.

\item Finally, the total flux density of Io, $F_{\rm{Io}}\,[\rm{Jy}]$, was converted into W/$\mathrm{m}^{2}$/ $\mu$m and the normalized DIP images were multiplied by this value to obtain W/$\mathrm{m}^{2}$/ $\mu$m per pixel across the moon's disk. 
\end{itemize}

With the calibrated images, we computed Io's total spectral radiance (GW/$\mu$m) per pixel, considering a distance of 4.363 au between the moon and JWST at the time of the observations. The total radiance of Io is I$_{Io}\, \sim$ 2450 GW/$\mu$m. Considering that the resolution of NIRISS-AMI ($\theta_{\lambda/\mathrm{2B}}\, \sim$  84 mas) is close to the scale of the pixel size ( $\sim$ 65.7 mas), we adopted the spectral radiance of each one of the volcanoes by simply determining the radiance at the peak of the emission at each volcano. This is a lower limit to the actual radiance, as explained by, e.g., \citet{DEPATER_2014}. From our experiments, the reported values are close to half of the total radiance in the volcanoes (see Appendix \ref{sec:disk_sim_flux}). To fully characterize the total radiance from each volcano, we need to take into account the background emission (e.g., scattering, extended emission, etc), crowding, and the response of the synthesized interferometric beam. However, due to the coarse sampling of the AMI data, we decided to report the radiance at the peak of each volcano. Using the positions of the hot spots as summarized in Table \ref{tab:volcanoes_fluxes}, we corrected the observed radiances by dividing the numbers by the cosine of the emission angle. The mean values and their standard deviations reported in Table \ref{tab:volcanoes_fluxes} were calculated by averaging the volcanoes' radiances obtained at each one of the DIP deconvolved images of the 100 integrations per science exposure. A final average (over the five science exposures in our data) and its 1-$\sigma$ uncertainty is also reported per volcano.



\begin{table*}
\caption{Volcanoes identified in the AMI data and their spectral radiances}
\begin{threeparttable}[b]
\label{tab:volcanoes_fluxes}
\begin{tabular}{|| l c c c c c c c||} 
\hline

& & \multicolumn{5} {c|}{Radiances [GW/$\mu$m] per epoch$^{1}$} \\
\cline{3-7}
Volcano name & Coordinates [Lon., Lat.] & 16:32:23$^{2}$ & 16:40:56 & 16:49:29 & 16:58:02 & 17:06:35 & Average\\
 \hline
  North of Seth Patera (V1)  & 129.4 $\pm$ 0.8$^{\circ}$ W, 1.5 $\pm$ 0.7$^{\circ}$ S        & 29 $\pm$ 1.6   & 35 $\pm$ 3.2   & 33 $\pm$ 1.4   & 31 $\pm$ 1.7   & 35 $\pm$ 1.0   & 33 $\pm$ 4.3 \\
  P 197 (V2)                 & 105.8 $\pm$ 1.2$^{\circ}$ W, 45.0 $\pm$ 1$^{\circ}$ S         & 20 $\pm$ 1.3   & 24 $\pm$ 2.1   & 25 $\pm$ 1.2   & 22 $\pm$ 1.4   & 26 $\pm$ 1.8   & 23 $\pm$ 3.6 \\  
  Masubi (V3)               & 50.0 $\pm$ 1$^{\circ}$ W, 44.0 $\pm$ 0.3$^{\circ}$ S          & 38 $\pm$ 3.9    & 41 $\pm$ 7.5   & 33 $\pm$ 3.1   & 15 $\pm$ 1.0   & 24 $\pm$ 4.2   & 30 $\pm$ 10.0 \\
  N. Lei-Zi Fluctus (V4)     & 42.8 $\pm$ 0.8$^{\circ}$ W, 24.2 $\pm$ 1.4$^{\circ}$ N        & 34 $\pm$ 5.5   & 36 $\pm$ 6.4   & 20 $\pm$ 1.5   & 33 $\pm$ 2.9   & 27 $\pm$ 1.2   & 30 $\pm$ 9.1 \\
  Kanehekili Fluctus (V5)    & 47.6 $\pm$ 1.5$^{\circ}$ W, 17.4 $\pm$ 1.2$^{\circ}$ S        & 13 $\pm$ 1 3   & 18 $\pm$ 2.3   & 14 $\pm$ 1.9   & 13 $\pm$ 0.5   & 15 $\pm$ 2.2   & 14 $\pm$ 4.0 \\
     Amirani (V6)              & 110.6 $\pm$ 1.3$^{\circ}$ W, 22.3 $\pm$ 2$^{\circ}$ N          & 11 $\pm$ 0.3   & 11 $\pm$ 0.6   & 11 $\pm$ 0.4   & 10 $\pm$ 0.3   & 11 $\pm$ 0.3   & 11 $\pm$ 0.8 \\
    Tonatiuh (V7)               & 80.2 $\pm$ 0.7$^{\circ}$ W, 53.3 $\pm$ 1.3$^{\circ}$ N     & 17 $\pm$ 1.0   & 17 $\pm$ 1.0   & 13 $\pm$ 0.4   & 18 $\pm$ 1.2   & 16 $\pm$ 1.9   & 16 $\pm$ 2.7 \\

 \hline
\end{tabular}
\begin{tablenotes}
       \item [1] The reported radiances at the peak of the emission correspond to approximately half of the total flux
       \item [2] Universal Time on August 1st, 2022
     \end{tablenotes}
 \end{threeparttable}
\end{table*}

\subsection{The size of  the structures on Io's disk}

As shown, the brightest structures on the surface of Io are hot spots. However, we also observe changes in the brightness of the disk's background. Those changes are also seen in the Keck images, where some extended emission is observed around V1 (see for example the first panel in Fig.\ref{fig:keckIo}). In order to characterize the dominant sizes of those observed structures, we computed the power spectral density, PSD(k), of the object. For this, we used the median DIP images and we employed the method described by \citet{Rosales-Guzman_2024}. First, to remove the median stellar disk contribution from the PSD, we estimated the median pixel value across the moon's disk, and we defined this value as the minimum one for all the pixels across the moon's disk. Second, we computed the Discrete Fourier Transform (DFT) of the images considering the inverse of the pixel scale (1/65.7 cycles/mas) as the maximum spatial frequency sampled in the DFT. Finally, we calculated the momentum of the power spectrum, P(k), per frequency, k: PSD(k) = k × P(k). 

\begin{figure}
\centering
    \includegraphics[width=0.96\linewidth]{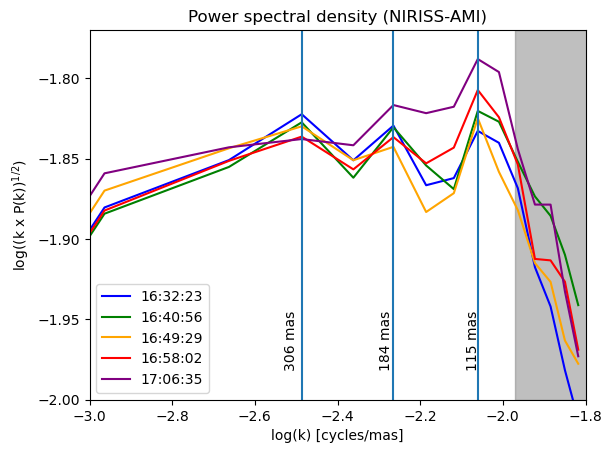}
    \includegraphics[width=0.95\linewidth]{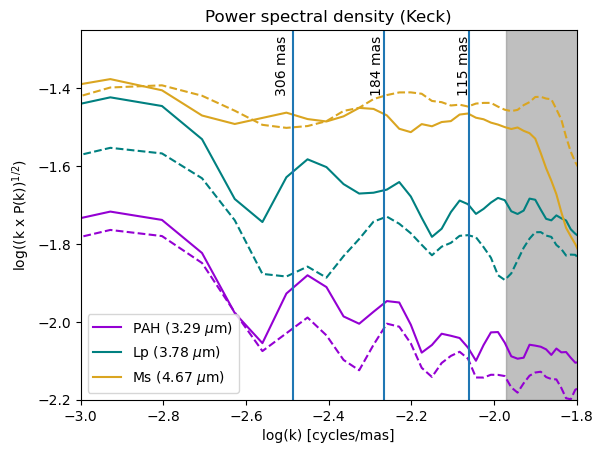}
    \caption{Radial averaged power spectra of the deconvolved images (top panel) and of the Keck images (bottom panel). The angular scales traced by the three main peaks in the NIRISS-AMI data are shown with vertical blue lines in both panels. The colored continuous lines in the top panel indicates the PSD(k) of the five DIP deconvolved images. In the bottom panel, the continuous lines indicate the PSD(k) of the Keck images taken in July, 2022, while the dashed lines shows the PSD(k) of the Keck data in September, 2022. The different colors indicate different filters. The gray shaded area corresponds to spatial frequencies higher than the best resolution element of the AMI interferometer.}
    \label{fig:psd}
\end{figure}

The upper panel in Figure \ref{fig:psd} displays the PSD(k) obtained from the science DIP mean images. There are three main peaks well-defined and coincident in all the data sets. Considering that the angular diameter of Io's disk at the time of the observations is 1.156 arcsec, and that the radius at the equator is $R_{\mathrm{Io}}$ = 1821.6 km, those peaks trace scales of 362 km (115 mas), 580 km (184 mas) and 964 km (306 mas), which is much larger than the smearing induced by Io's rotation ($\sim$26 km at the center of Io's disk). In all the epochs the highest peak traces a scale of 362 km that, roughly, corresponds to the sizes of emission region at the individual volcanoes at this wavelength. The peak that traces the 580 km scale may be related to the presence of SO$_2$ frost near the equator as mapped in the SO$_2$-ice bands at 1.98 $\mu$m and 2.12 $\mu$m \citep{laver2008,Laver_2009, depater2020IoSO}. Notice that the extended emission near V1 is present in both the JWST-AMI and Keck images (more visible at 3.29 $\mu$m and 4.78 $\mu$m). 

For comparison, we also used the method described above with the normalized Keck images of the two observed epochs. The lower panel in Fig. \ref{fig:psd} shows the PSD(k) obtained from those images. The PSD(k) of the Keck's PAH and Lp filters from July 9, 2022 show more peaks than the three ones in the NIRISS-AMI data. This suggests that there are more dominant structures at different angular scales across the moon's disk. However, when we compare the PSD(k) from Keck's Ms filter, we see that the three distinct NIRISS-AMI peaks appear. Nevertheless, they are not exactly coincident with the ones observed in the JWST PSD(k). Instead they reach their maxima at 119, 208 and 316 mas which, at a distance of d$^{\mathrm{Aug}}_{\mathrm{Ms}}$ = 4.695 au, correspond to 404, 705 and 1079 km. The PAH and Lp filters show peaks at 128, 169, and 283 mas, which traces 434, 573 and 960 km, respectively. Those values are closer to the ones obtained from the NIRISS-AMI data.

We identify three main causes of the differences observed between the NIRISS-AMI and Keck PSD(k). The first one is the difference in wavelength and resolution. The Keck PAH filter is the one with the closest resolution to the NIRISS-AMI one. However, at the same time, it is the one with the largest difference in wavelength. Second, we notice that the first diffraction ring in Keck's Ms filter is very prominent in the image. Hence, the AO quasi-static speckles around the diffraction ring can affect the correct characterization of the extended emission around it. Third, the NRM acts as spatial filter, thus it reduces the flux contribution of the structures that departs in size from the spatial scales traced by the different baselines. This is very effective to highlight particular angular scales. In the case of the NIRISS-AMI data, it helps us to better distinguish the structures across the Io's disk, despite of having a very coarse sampling at the detector. For our 4.3 $\mu$m data, a resolution element corresponds to 1.28 pixels, which is below the Nyquist sampling limit. This is not the case of the Keck data, which have finer sampling (8.5 pixels correspond to one resolution element for the PAH filter). At the same time, the direct AO imaging samples all the spatial scales, which helps to map the entire disk. However, it is more difficult to remove the background when computing the PSD(k). This effect shows up at Keck's PSD(k) since the biggest peak traces scales of 845 mas ($\sim$ 2866 km). This angular scale is indeed larger than the scale traced by the shortest baseline in the NRM (at $\lambda$/B, it corresponds to 670 mas or $\sim$ 2111 km).

When we compare the  PSD(k) from Keck's September epoch with the NIRISS-AMI one, we observe that it differs even more than the July epoch. There are several causes for this. The first one is that the distance is significantly closer, d$^{\mathrm{Sep.}}_{\mathrm{Ms}}$ = 3.983 au, compared with Keck's July epoch and the NIRISS-AMI one. Therefore, the projected disk diameter is of 1267.6 mas, and the peaks in the PSD(k) trace different linear scales compared with the other two epochs. Second, when we observe some of the identified volcanoes (V1, V2 and V6) in the September epoch, we notice that close to V2 there is an adjacent bright spot that overlaps with it, which effectively corresponds to a different angular scale in the PSD(k). These changes clearly illustrate the dynamic environment of Io. In the course of only two months, the brightness, position and number of volcanoes change considerably across the moon's disk.

\begin{figure*}
    \includegraphics[width=13 cm]{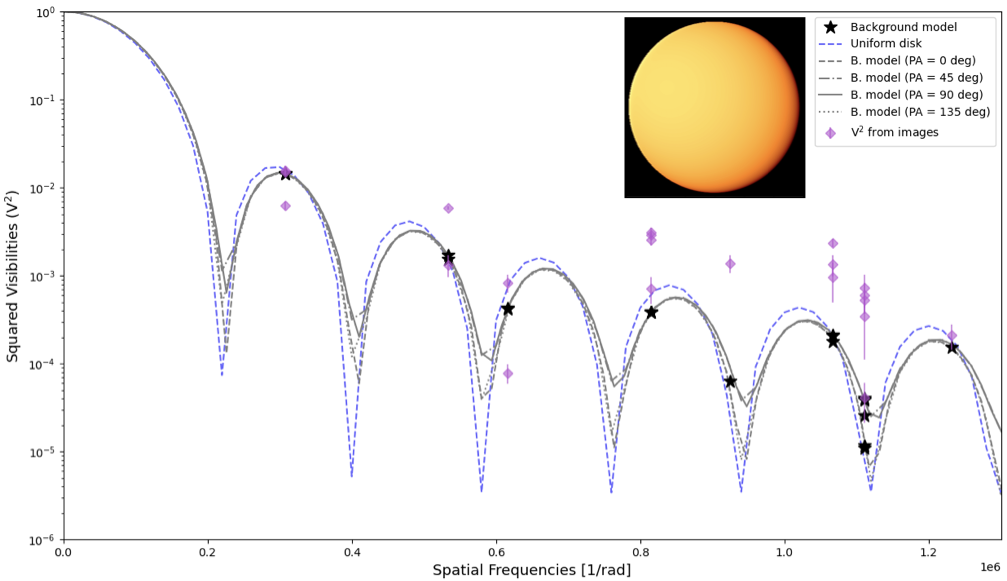}
    \caption{Squared visibilities versus spatial frequencies. The violet diamonds correspond to the mean V$^2$ extracted from DIP images. The black stars correspond to the V$^2$ extracted from the Minnaert background model at the baselines traced with the NRM. The gray lines trace the visibility function of the Minnaert model at different position angles and the blue dashed-line displays the visibility function of a uniformly illuminated disk model. The inset (top-left) displays the Minnaert illumination model with $k$ = 0.65. }
    \label{fig:vis}
\end{figure*}

While the PSD(k) offers us an effective way to determine the sizes of the structures that contribute more to the observed emission directly from the deconvolved images, we have already established that the NRM acts as a spatial filter. Thus, as an additional method to characterize the structures of the moon's surface, it is also interesting to determine the distribution of the emission over the baselines of the NRM. For this, we computed the squared visibilities (V$^2$) from the DIP deconvolved images. Fig \ref{fig:vis} displays the mean V$^2$ and their standard deviation from the five DIP deconvolved images of the AMI science exposures. For comparison, we plotted the visibility curve of a uniformly illuminated disk and four different visibility curves (with u-v sampling at different position angles) of a non-uniform background model of Io. The background model consist of an illumination map of Io at the time of the observations, that follows a Minnaert's law \citep[see e.g.,][]{Veverka_1989} which describes sunlight scattered of Io's surface:  

\begin{equation}
I/F = B_0 \mu_0^k \mu^{k-1}\,,
\end{equation}

where $B_0$ is similar to the normal albedo of the surface and $k$ is the limb-darkening coefficient. The cosine of the incidence angle is denoted by $\mu_0$ and the cosine of the emission angle by $\mu$.  

The visibility curves show that the NRM is tracing scales between the second and seventh lobes of a projected Minnaert's illuminated disk (or of a uniform disk)  with a diameter of 1156 mas. For the  five longest sets of baselines, we determine that the group of visibilites from the DIP images are above the predicted visibility value of the background model. This is expected since those baselines map scales between 167 and 253 mas. Notice that these scales are quite similar to the ones traced by the three main peaks in the PSD(k). The three shortest sets of baselines, tracing scales between 335 and 670 mas, show several visibilities below the expected values of the background model. We identify that the position angle of the baselines that produce these visibilities ranges between -112$^{\circ}$ and -172$^{\circ}$ East of North. At these position angles, the disk structure is fainter compared with other parts of the disk (except perhaps at the edges). While the Minnaert's background model offers a good proxy of the moon's scattered light, from the AMI and Keck images we observe that the surface of the moon is quite complex, therefore, it is not surprising to have areas that look fainter than the simple background model used here. Additionally, we used $k = 0.65$ for the limb-darkening coefficient. This value is an average over the surface of the disk and it can vary in different directions. Small changes in this parameter (for example $k = 0.8$) causes noticeable differences in the visibility curve.           

\subsection{V1/Seth Patera motion}

\begin{figure*}
    \includegraphics[width=16 cm]{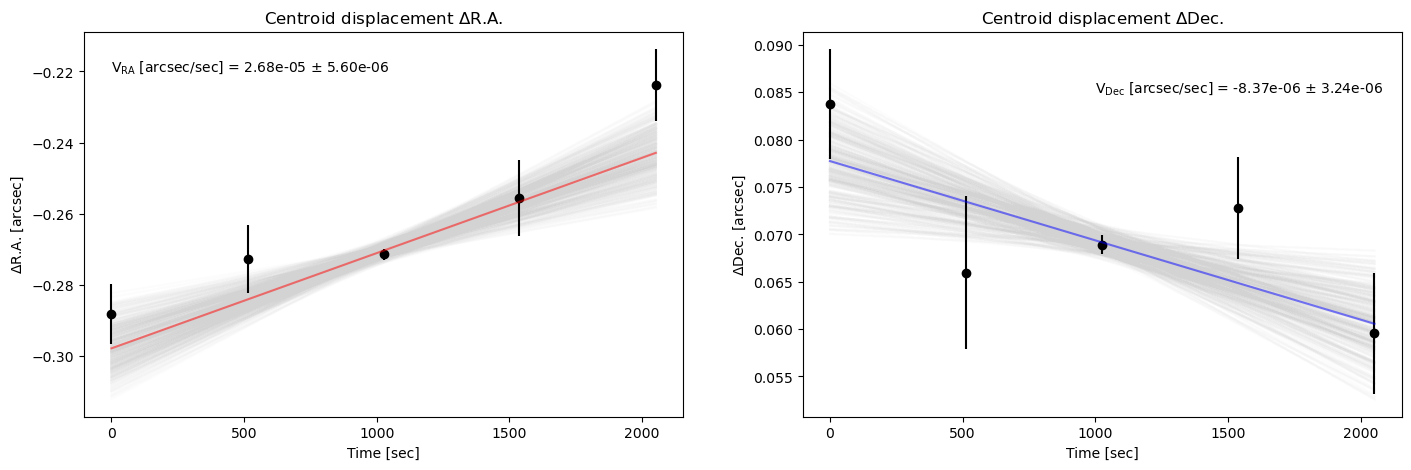}
    \caption{Proper motions of the brightness component, V1, in the surface of Io. The left panel displays the Right Ascension's flux centroid of V1. Similarly, the right panel displays the Declination's flux centroid of V1. The best-fit model is displayed with red and blue dashed-lines in R.A and Dec., respectively. The gray lines indicate other fit solutions within 1$\sigma$ of the mean value.}
    \label{fig:centroid_mass}
\end{figure*}

Due to the rotation of Io, we observed a small displacement (to the right of the images) of the hot spots across the five DIP images. We used the brightest feature, V1, to characterize this motion. For this, we computed the flux centroid of this feature using each one of the integrations in the deconvolved science exposures. First, we de-projected the deconvolved DIP integrations in a Lon.-Lat. map. Second, we computed the longitude and latitude of the flux centroid of V1 using a circular mask with a radius of four pixels around the brightest pixel of V1. The extraction of the centroid was obtained with the function \texttt{center\_of\_mass} inside the \texttt{Python} package \texttt{scipy.ndimage}. Third, we converted the longitude and latitude of the flux centroid into relative angular coordinates (in $\Delta$R.A. and $\Delta$Dec.) from the center of the disk. Finally, the flux centroids were averaged over each one of the exposures. Figure \ref{fig:centroid_mass} displays the mean centroid positions relative to the center of the moon per data cube and their corresponding standard deviation. 

We fitted a straight line to the R.A. and Dec. components versus time, which allowed us to compute the observed speed, using the \texttt{Python} package \texttt{emcee}. It employs a Monte Carlo - Markov Chain method to explore the parameter space. We used 5000 independent chains that evolved over 1000 steps. The best-fit model provides the following projected components of the V1 velocity: V$_{\mathrm{RA}}$ = -2.7 $\pm$ 0.6  $\times$ 10$^{-5}$ arcsec s$^{-1}$ and V$_{\mathrm{Dec}}$ = 8.4 $\pm$ 3.2  $\times$ 10$^{-6}$ arcsec s$^{-1}$, which is equivalent to a transverse speed of  V$_{\mathrm{T}}$ = 86 $\pm$ 34 m s$^{-1}$. This speed is equal to the rotation speed at the equator V$_{\mathrm{rot}}$ = 75 m s$^{-1}$ (or V$_{\mathrm{rot}}$ = 271 km h$^{-1}$) within the uncertainties of the data.  This is a comforting sanity check on the retrieved positions; however, the uncertainties are still relative large. This could be caused by a combined effect of the sparse sampling of the deconvolved images and the extended emission around the peak of the emission of V1, which, certainly is affecting the flux centroid extraction. Additionally, the time difference between the first and last  science exposure is still small to properly trace the moon's changes across a rotational period (P$_{\mathrm{Io}}$ = 1.769 days), indeed, we only monitor $\sim$ 1\% of Io's rotation. Thus, a more complete monitoring of the source is necessary for a better characterization of the flux centroid of the volcanoes over time.

\section{Conclusion}

The analysis of JWST-NIRISS Aperture Masking Interferometric observations of Io demonstrates the critical role of novel deconvolution methods in enhancing our understanding of its volcanic activity. The use of neural networks for interferogram analysis has enabled a more detailed characterization of Io's surface, overcoming the limitations of traditional interferometric techniques. In particular, we observed that unsupervised NNs, like the Deep Image Prior algorithm used here, offer promising ways to recover realistic images compared with supervised NNs and other, more traditional, algorithms. The use of this diffusive autoencoder method allowed us to characterize the surface of Io at different scales, revealing key volcanic loci and their brightness, consistently confirmed with ground-based Keck AO imaging. 

Specifically, the non-redundant mask observations were instrumental in providing high-resolution data, overcoming the limitations of traditional interferometric techniques. The recovered images reveal complex structures on the surface of Io, from hot spots to extended emission, which can be associated with volcanic activity on this Jovian satellite. A more consistent analysis of the temporal evolution of the hot spots and extended emission requires a more complete monitoring of the surface with the highest angular resolution possible. For this, JWST-NIRISS AMI and other ground-based facilities (like e.g., Keck and the Large Binocular Telescope Interferometer) play a key role. 

\section*{Acknowledgments}

The authors acknowledge the comments and suggestions made by the referee to improve the manuscript. This work is based on observations made with the NASA/ESA/CSA James Webb Space Telescope. The data were obtained from the Mikulski Archive for Space Telescopes at the Space Telescope Science Institute, which is operated by the Association of Universities for Research in Astronomy, Inc., under NASA contract NAS 5-03127 for JWST. The observations are associated with the Early Release Science (ERS) program \#1373, which is led by co-PIs Imke de Pater and Thierry Fouchet.

J.S.-B. acknowledges the support received by the UNAM DGAPA-PAPIIT projects IA 105023 and AG 101025. J.S.-B. thanks J. K. Barrera-Ballesteros for the fruitful discussions about the volcanoes' radiance calculation. 

Data presented in this paper were also obtained at the W.M. Keck
Observatory, which is operated as a scientific partnership among the California Institute of Technology, the University
of California, and the National Aeronautics and Space
Administration. The Observatory was made possible by the
generous financial support of the W.M. Keck Foundation.
The authors wish to recognize and acknowledge the very
significant cultural role and reverence that the summit of
Maunakea has always had within the indigenous Hawaiian
community. Without their generous hospitality,
we would not have been able to obtain the observations presented here.

We thank the Keck Telescope Operators and Support Astronomers Carlos Alvarez and Randy Campbell for their continued support of the Twilight Zone Project. The data on July 9, 2022, were taken under the Twilight Zone Project by A. Connors (see \url{https://www2.keck.hawaii.edu/inst/tda/TwilightZone.html#}).

We thank John Stansberry for initiating and supporting early feasibility studies of these observations.

IdP is in part supported by the Space Telescope Science Institute grant nr. JWST-ERS-01373.

{\bf Author Contributions:} JSB reduced and analyzed the NIRISS-AMI data and wrote the first draft of the paper. 
IdP and TF are the Co-PIs of the DD-ERS proposal on the Jovian system that obtained the data. 
IdP reduced and analyzed the Keck data, constructed Figures \ref{fig:keckIo} and \ref{fig:USGS}, and wrote parts of some sections.
While a visiting scholar at the keck Observatory, EMM wrote the observing scripts for  the Twilight Zone observing campaign and provided models of Io's infrared flux.
AS and DT conducted a feasiblity study of the observations, and RC provided JWST pipeline data analysis and code. All authors read and gave valuable comments on the paper.
\section*{Data Availability}

The raw and processed AMI data underlying this paper will be shared on
reasonable request to the corresponding author.
All Keck data are public on the Keck Archive (\url{https://koa.ipac.caltech.edu/cgi-bin/KOA/nph-KOAlogin}.



\bibliographystyle{mnras}
\bibliography{example} 

\begin{thebibliography}{}
\makeatletter
\relax
\def\mn@urlcharsother{\let\do\@makeother \do\$\do\&\do\#\do\^\do\_\do\%\do\~}
\def\mn@doi{\begingroup\mn@urlcharsother \@ifnextchar [ {\mn@doi@}
  {\mn@doi@[]}}
\def\mn@doi@[#1]#2{\def\@tempa{#1}\ifx\@tempa\@empty \href
  {http://dx.doi.org/#2} {doi:#2}\else \href {http://dx.doi.org/#2} {#1}\fi
  \endgroup}
\def\mn@eprint#1#2{\mn@eprint@#1:#2::\@nil}
\def\mn@eprint@arXiv#1{\href {http://arxiv.org/abs/#1} {{\tt arXiv:#1}}}
\def\mn@eprint@dblp#1{\href {http://dblp.uni-trier.de/rec/bibtex/#1.xml}
  {dblp:#1}}
\def\mn@eprint@#1:#2:#3:#4\@nil{\def\@tempa {#1}\def\@tempb {#2}\def\@tempc
  {#3}\ifx \@tempc \@empty \let \@tempc \@tempb \let \@tempb \@tempa \fi \ifx
  \@tempb \@empty \def\@tempb {arXiv}\fi \@ifundefined
  {mn@eprint@\@tempb}{\@tempb:\@tempc}{\expandafter \expandafter \csname
  mn@eprint@\@tempb\endcsname \expandafter{\@tempc}}}

\bibitem[\protect\citeauthoryear{{Artigau} et~al.,}{{Artigau}
  et~al.}{2014}]{Artigau_2014}
{Artigau} {\'E}.,  et~al., 2014, in {Oschmann} Jacobus~M. J.,  {Clampin} M.,
  {Fazio} G.~G.,   {MacEwen} H.~A.,  eds,  Society of Photo-Optical
  Instrumentation Engineers (SPIE) Conference Series Vol. 9143, Space
  Telescopes and Instrumentation 2014: Optical, Infrared, and Millimeter Wave.
  p. 914340 (\mn@eprint {arXiv} {1406.6882}), \mn@doi{10.1117/12.2055191}

\bibitem[\protect\citeauthoryear{Bishop}{Bishop}{1995}]{Bishop_1995}
Bishop C.~M.,  1995, Neural networks for pattern recognition.
Oxford university press

\bibitem[\protect\citeauthoryear{{Blakely} et~al.,}{{Blakely}
  et~al.}{2022}]{Blakely_2022}
{Blakely} D.,  et~al., 2022, \mn@doi [\apj] {10.3847/1538-4357/ac6586}, \href
  {https://ui.adsabs.harvard.edu/abs/2022ApJ...931....3B} {931, 3}

\bibitem[\protect\citeauthoryear{{Blakely} et~al.,}{{Blakely}
  et~al.}{2025}]{Blakely_2025}
{Blakely} D.,  et~al., 2025, \mn@doi [\aj] {10.3847/1538-3881/ad9b94}, \href
  {https://ui.adsabs.harvard.edu/abs/2025AJ....169..137B} {169, 137}

\bibitem[\protect\citeauthoryear{{Cheetham}, {Girard}, {Lacour}, {Schworer},
  {Haubois}  \& {Beuzit}}{{Cheetham} et~al.}{2016}]{Cheetham_2016}
{Cheetham} A.~C.,  {Girard} J.,  {Lacour} S.,  {Schworer} G.,  {Haubois} X.,
  {Beuzit} J.-L.,  2016, in {Malbet} F.,  {Creech-Eakman} M.~J.,   {Tuthill}
  P.~G.,  eds,  Society of Photo-Optical Instrumentation Engineers (SPIE)
  Conference Series Vol. 9907, Optical and Infrared Interferometry and Imaging
  V. p. 99072T, \mn@doi{10.1117/12.2231983}

\bibitem[\protect\citeauthoryear{Davies}{Davies}{2007}]{Davies_2007}
Davies A.~G.,  2007, Volcanism on Io: A Comparison with Earth.
Cambridge Planetary Science, Cambridge University Press

\bibitem[\protect\citeauthoryear{{Davies}, {Perry}, {Williams}, {Veeder}  \&
  {Nelson}}{{Davies} et~al.}{2024}]{davies2024}
{Davies} A.~G.,  {Perry} J.~E.,  {Williams} D.~A.,  {Veeder} G.~J.,   {Nelson}
  D.~M.,  2024, \mn@doi [The Planetary Science Journal] {10.3847/PSJ/ad4346},
  \href {https://ui.adsabs.harvard.edu/abs/2024PSJ.....5..121D} {5, 121}

\bibitem[\protect\citeauthoryear{{Doyon} et~al.,}{{Doyon}
  et~al.}{2023}]{Doyon_2023}
{Doyon} R.,  et~al., 2023, \mn@doi [\pasp] {10.1088/1538-3873/acd41b}, \href
  {https://ui.adsabs.harvard.edu/abs/2023PASP..135i8001D} {135, 098001}

\bibitem[\protect\citeauthoryear{Egmont-Petersen, {de Ridder}  \&
  Handels}{Egmont-Petersen et~al.}{2002}]{Egmont-Petersen_2002}
Egmont-Petersen M.,  {de Ridder} D.,   Handels H.,  2002, \mn@doi [Pattern
  Recognition] {https://doi.org/10.1016/S0031-3203(01)00178-9}, 35, 2279

\bibitem[\protect\citeauthoryear{Flamary}{Flamary}{2017}]{Flamary_2017}
Flamary R.,  2017, in 2017 25th European Signal Processing Conference
  (EUSIPCO). pp 2468--2472, \mn@doi{10.23919/EUSIPCO.2017.8081654}

\bibitem[\protect\citeauthoryear{{Gallenne}, {Desgrange}, {Milli},
  {Sanchez-Bermudez}, {Chauvin}, {Kraus}, {Girard}  \& {Boccaletti}}{{Gallenne}
  et~al.}{2022}]{Gallenne_2022}
{Gallenne} A.,  {Desgrange} C.,  {Milli} J.,  {Sanchez-Bermudez} J.,  {Chauvin}
  G.,  {Kraus} S.,  {Girard} J.~H.,   {Boccaletti} A.,  2022, \mn@doi [\aap]
  {10.1051/0004-6361/202244226}, \href
  {https://ui.adsabs.harvard.edu/abs/2022A&A...665A..41G} {665, A41}

\bibitem[\protect\citeauthoryear{{Greenbaum}, {Pueyo}, {Sivaramakrishnan}  \&
  {Lacour}}{{Greenbaum} et~al.}{2015}]{Greenbaum_2015}
{Greenbaum} A.~Z.,  {Pueyo} L.,  {Sivaramakrishnan} A.,   {Lacour} S.,  2015,
  \mn@doi [\apj] {10.1088/0004-637X/798/2/68}, \href
  {https://ui.adsabs.harvard.edu/abs/2015ApJ...798...68G} {798, 68}

\bibitem[\protect\citeauthoryear{{Hubel} \& {Wiesel}}{{Hubel} \&
  {Wiesel}}{1959}]{Hubel_Wiesel_1959}
{Hubel} D.,  {Wiesel} T.,  1959, J. Physiol., 148, 574–591

\bibitem[\protect\citeauthoryear{{Isbell} et~al.,}{{Isbell}
  et~al.}{2022}]{Isbell_2022}
{Isbell} J.~W.,  et~al., 2022, \mn@doi [\aap] {10.1051/0004-6361/202243271},
  \href {https://ui.adsabs.harvard.edu/abs/2022A&A...663A..35I} {663, A35}

\bibitem[\protect\citeauthoryear{Jamal \& Bloom}{Jamal \&
  Bloom}{2020}]{Jamal_2020}
Jamal S.,  Bloom J.~S.,  2020, \mn@doi [The Astrophysical Journal Supplement
  Series] {10.3847/1538-4365/aba8ff}, 250, 30

\bibitem[\protect\citeauthoryear{{Lacour}, {Perrin}, {Woillez}, {Assemat}  \&
  {Thiebault}}{{Lacour} et~al.}{2004}]{Lacour_2004}
{Lacour} S.,  {Perrin} G.~S.,  {Woillez} J.~M.,  {Assemat} F.,   {Thiebault}
  E.~M.,  2004, in {Traub} W.~A.,  ed.,  Society of Photo-Optical
  Instrumentation Engineers (SPIE) Conference Series Vol. 5491, New Frontiers
  in Stellar Interferometry. p.~1342, \mn@doi{10.1117/12.551746}

\bibitem[\protect\citeauthoryear{{Lau} et~al.,}{{Lau} et~al.}{2024}]{Lau_2024}
{Lau} R.~M.,  et~al., 2024, \mn@doi [\apj] {10.3847/1538-4357/ad192c}, \href
  {https://ui.adsabs.harvard.edu/abs/2024ApJ...963..127L} {963, 127}

\bibitem[\protect\citeauthoryear{{Laver} \& {de Pater}}{{Laver} \& {de
  Pater}}{2008}]{laver2008}
{Laver} C.,  {de Pater} I.,  2008, \mn@doi [\icarus]
  {10.1016/j.icarus.2007.12.023}, \href
  {https://ui.adsabs.harvard.edu/abs/2008Icar..195..752L} {195, 752}

\bibitem[\protect\citeauthoryear{{Laver} \& {de Pater}}{{Laver} \& {de
  Pater}}{2009}]{Laver_2009}
{Laver} C.,  {de Pater} I.,  2009, \mn@doi [\icarus]
  {10.1016/j.icarus.2008.12.037}, \href
  {https://ui.adsabs.harvard.edu/abs/2009Icar..201..172L} {201, 172}

\bibitem[\protect\citeauthoryear{Lecun, Boser, Denker, Henderson, Howard,
  Hubbard  \& Jackel}{Lecun et~al.}{1990}]{Lecun_1980}
Lecun Y.,  Boser B.,  Denker J.,  Henderson D.,  Howard R.,  Hubbard W.,
  Jackel L.,  1990, in Touretzky D.,  ed.,  Vol. 2, Advances in Neural
  Information Processing Systems (NIPS 1989), Denver, CO. Morgan Kaufmann

\bibitem[\protect\citeauthoryear{{Leggett} et~al.,}{{Leggett}
  et~al.}{2003}]{leggett2003}
{Leggett} S.~K.,  et~al., 2003, \mn@doi [\mnras]
  {10.1046/j.1365-8711.2003.06943.x}, \href
  {https://ui.adsabs.harvard.edu/abs/2003MNRAS.345..144L} {345, 144}

\bibitem[\protect\citeauthoryear{{Lopes} \& {Spencer}}{{Lopes} \&
  {Spencer}}{2007}]{Lopes_2007}
{Lopes} R. M.~C.,  {Spencer} J.~R.,  2007, {Io After Galileo: A New View of
  Jupiter's Volcanic Moon}, \mn@doi{10.1007/978-3-540-48841-5.
}

\bibitem[\protect\citeauthoryear{{Lucy}}{{Lucy}}{1974}]{Lucy_1974}
{Lucy} L.~B.,  1974, \mn@doi [\aj] {10.1086/111605}, \href
  {https://ui.adsabs.harvard.edu/abs/1974AJ.....79..745L} {79, 745}

\bibitem[\protect\citeauthoryear{{Marchis} et~al.,}{{Marchis}
  et~al.}{2002}]{marchis2002}
{Marchis} F.,  et~al., 2002, \mn@doi [\icarus] {10.1006/icar.2002.6955}, \href
  {https://ui.adsabs.harvard.edu/abs/2002Icar..160..124M} {160, 124}

\bibitem[\protect\citeauthoryear{Marr \& Hildreth}{Marr \&
  Hildreth}{1980}]{Marr_1980}
Marr D.,  Hildreth E.,  1980, Proceedings of the Royal Society of London.
  Series B, Biological Sciences, 207, 187

\bibitem[\protect\citeauthoryear{{Matson}, {Ransford}  \& {Johnson}}{{Matson}
  et~al.}{1981}]{Matson_1981}
{Matson} D.~L.,  {Ransford} G.~A.,   {Johnson} T.~V.,  1981, \mn@doi [\jgr]
  {10.1029/JB086iB03p01664}, \href
  {https://ui.adsabs.harvard.edu/abs/1981JGR....86.1664M} {86, 1664}

\bibitem[\protect\citeauthoryear{{Morabito}, {Synnott}, {Kupferman}  \&
  {Collins}}{{Morabito} et~al.}{1979}]{Morabito_1979}
{Morabito} L.~A.,  {Synnott} S.~P.,  {Kupferman} P.~N.,   {Collins} S.~A.,
  1979, \mn@doi [Science] {10.1126/science.204.4396.972}, \href
  {https://ui.adsabs.harvard.edu/abs/1979Sci...204..972M} {204, 972}

\bibitem[\protect\citeauthoryear{Narayan \& Nityananda}{Narayan \&
  Nityananda}{1986}]{Narayan_1986}
Narayan R.,  Nityananda R.,  1986, \mn@doi [Annual Review of Astronomy and
  Astrophysics] {https://doi.org/10.1146/annurev.aa.24.090186.001015}, 24, 127

\bibitem[\protect\citeauthoryear{{Peale}, {Cassen}  \& {Reynolds}}{{Peale}
  et~al.}{1979}]{Peale_1979}
{Peale} S.~J.,  {Cassen} P.,   {Reynolds} R.~T.,  1979, \mn@doi [Science]
  {10.1126/science.203.4383.892}, \href
  {https://ui.adsabs.harvard.edu/abs/1979Sci...203..892P} {203, 892}

\bibitem[\protect\citeauthoryear{Perry, Davies, Williams  \& Nelson}{Perry
  et~al.}{2025}]{Perry_2025}
Perry J.~E.,  Davies A.~G.,  Williams D.~A.,   Nelson D.~M.,  2025, \mn@doi
  [The Planetary Science Journal] {10.3847/PSJ/adbae3}, 6, 84

\bibitem[\protect\citeauthoryear{Rathbun \& Spencer}{Rathbun \&
  Spencer}{2006}]{Rathbun_2006}
Rathbun J.~A.,  Spencer J.~R.,  2006, \mn@doi [Geophysical Research Letters]
  {https://doi.org/10.1029/2006GL026844}, 33

\bibitem[\protect\citeauthoryear{{Rathbun}, {Spencer}, {Lopes}  \&
  {Howell}}{{Rathbun} et~al.}{2014}]{Rathbun_2014}
{Rathbun} J.~A.,  {Spencer} J.~R.,  {Lopes} R.~M.,   {Howell} R.~R.,  2014,
  \mn@doi [\icarus] {10.1016/j.icarus.2013.12.002}, \href
  {https://ui.adsabs.harvard.edu/abs/2014Icar..231..261R} {231, 261}

\bibitem[\protect\citeauthoryear{{Redwing}, {de Pater}, {Luszcz-Cook}, {de
  Kleer}, {Moullet}  \& {Rojo}}{{Redwing} et~al.}{2022}]{redwing2022}
{Redwing} E.,  {de Pater} I.,  {Luszcz-Cook} S.,  {de Kleer} K.,  {Moullet} A.,
    {Rojo} P.~M.,  2022, \mn@doi [The Planetary Science Journal]
  {10.3847/PSJ/ac9784}, \href
  {https://ui.adsabs.harvard.edu/abs/2022PSJ.....3..238R} {3, 238}

\bibitem[\protect\citeauthoryear{Ren, Zhang, Wang, Hu  \& Zuo}{Ren
  et~al.}{2019}]{Ren_2019}
Ren D.,  Zhang K.,  Wang Q.,  Hu Q.,   Zuo W.,  2019, 2020 IEEE/CVF Conference
  on Computer Vision and Pattern Recognition (CVPR), pp 3338--3347

\bibitem[\protect\citeauthoryear{{Roe}, {de Pater}, {Macintosh}, {Gibbard},
  {Max}  \& {McKay}}{{Roe} et~al.}{2002}]{roe2002}
{Roe} H.~G.,  {de Pater} I.,  {Macintosh} B.~A.,  {Gibbard} S.~G.,  {Max}
  C.~E.,   {McKay} C.~P.,  2002, \mn@doi [\icarus] {10.1006/icar.2002.6831},
  \href {https://ui.adsabs.harvard.edu/abs/2002Icar..157..254R} {157, 254}

\bibitem[\protect\citeauthoryear{{Rosales-Guzm{\'a}n}
  et~al.,}{{Rosales-Guzm{\'a}n} et~al.}{2024}]{Rosales-Guzman_2024}
{Rosales-Guzm{\'a}n} A.,  et~al., 2024, \mn@doi [\aap]
  {10.1051/0004-6361/202349112}, \href
  {https://ui.adsabs.harvard.edu/abs/2024A&A...688A.124R} {688, A124}

\bibitem[\protect\citeauthoryear{{Sagan}}{{Sagan}}{1979}]{Sagan_1979}
{Sagan} C.,  1979, \mn@doi [\nat] {10.1038/280750a0}, \href
  {https://ui.adsabs.harvard.edu/abs/1979Natur.280..750S} {280, 750}

\bibitem[\protect\citeauthoryear{{Sallum} et~al.,}{{Sallum}
  et~al.}{2024}]{Sallum_2024}
{Sallum} S.,  et~al., 2024, \mn@doi [\apjl] {10.3847/2041-8213/ad21fb}, \href
  {https://ui.adsabs.harvard.edu/abs/2024ApJ...963L...2S} {963, L2}

\bibitem[\protect\citeauthoryear{{Sanchez-Bermudez}, {Sch{\"o}del}, {Alberdi},
  {Muzi{\'c}}, {Hummel}  \& {Pott}}{{Sanchez-Bermudez}
  et~al.}{2014}]{Sanchez-Bermudez_2014}
{Sanchez-Bermudez} J.,  {Sch{\"o}del} R.,  {Alberdi} A.,  {Muzi{\'c}} K.,
  {Hummel} C.~A.,   {Pott} J.~U.,  2014, \mn@doi [\aap]
  {10.1051/0004-6361/201423657}, \href
  {https://ui.adsabs.harvard.edu/abs/2014A&A...567A..21S} {567, A21}

\bibitem[\protect\citeauthoryear{{Sanchez-Bermudez}, {Alberdi}, {Sch{\"o}del}
  \& {Sivaramakrishnan}}{{Sanchez-Bermudez}
  et~al.}{2022}]{Sanchez-Bermudez_2022}
{Sanchez-Bermudez} J.,  {Alberdi} A.,  {Sch{\"o}del} R.,   {Sivaramakrishnan}
  A.,  2022, in {M{\'e}rand} A.,  {Sallum} S.,   {Sanchez-Bermudez} J.,  eds,
  Society of Photo-Optical Instrumentation Engineers (SPIE) Conference Series
  Vol. 12183, Optical and Infrared Interferometry and Imaging VIII. p. 121831K,
  \mn@doi{10.1117/12.2629488}

\bibitem[\protect\citeauthoryear{{Sanchez-Bermudez}, {Cruz-Osorio},
  {Barrera-Ballesteros}, {Alberdi}  \& {Sch{\"o}del}}{{Sanchez-Bermudez}
  et~al.}{2024}]{Sanchez-Bermudez_2024}
{Sanchez-Bermudez} J.,  {Cruz-Osorio} A.,  {Barrera-Ballesteros} J.~K.,
  {Alberdi} A.,   {Sch{\"o}del} R.,  2024, in {Kammerer} J.,  {Sallum} S.,
  {Sanchez-Bermudez} J.,  eds,  Society of Photo-Optical Instrumentation
  Engineers (SPIE) Conference Series Vol. 13095, Optical and Infrared
  Interferometry and Imaging IX. p. 1309519, \mn@doi{10.1117/12.3020320}

\bibitem[\protect\citeauthoryear{{Service}, {Lu}, {Campbell}, {Sitarski},
  {Ghez}  \& {Anderson}}{{Service} et~al.}{2016}]{service2016}
{Service} M.,  {Lu} J.~R.,  {Campbell} R.,  {Sitarski} B.~N.,  {Ghez} A.~M.,
  {Anderson} J.,  2016, \mn@doi [\pasp] {10.1088/1538-3873/128/967/095004},
  \href {https://ui.adsabs.harvard.edu/abs/2016PASP..128i5004S} {128, 095004}

\bibitem[\protect\citeauthoryear{{Sivaramakrishnan} et~al.,}{{Sivaramakrishnan}
  et~al.}{2010}]{2010SPIE.7731E..3WS}
{Sivaramakrishnan} A.,  et~al., 2010, in {Oschmann} Jr. J.~M.,  {Clampin}
  M.~C.,   {MacEwen} H.~A.,  eds,  Society of Photo-Optical Instrumentation
  Engineers (SPIE) Conference Series Vol. 7731, Space Telescopes and
  Instrumentation 2010: Optical, Infrared, and Millimeter Wave. p. 77313W,
  \mn@doi{10.1117/12.858161}

\bibitem[\protect\citeauthoryear{{Sivaramakrishnan} et~al.,}{{Sivaramakrishnan}
  et~al.}{2023}]{Sivaramakrishnan_2023}
{Sivaramakrishnan} A.,  et~al., 2023, \mn@doi [\pasp]
  {10.1088/1538-3873/acaebd}, \href
  {https://ui.adsabs.harvard.edu/abs/2023PASP..135a5003S} {135, 015003}

\bibitem[\protect\citeauthoryear{{Smith} et~al.,}{{Smith}
  et~al.}{1979}]{Smith_1979}
{Smith} B.~A.,  et~al., 1979, \mn@doi [Science] {10.1126/science.206.4421.927},
  \href {https://ui.adsabs.harvard.edu/abs/1979Sci...206..927S} {206, 927}

\bibitem[\protect\citeauthoryear{{Soulain} \& {Robert C.~M.~T.}}{{Soulain} \&
  {Robert C.~M.~T.}}{2023}]{Soulain_2023}
{Soulain} A.,  {Robert C.~M.~T.} 2023, {AMICAL: Aperture Masking Interferometry
  Calibration and Analysis Library}, Astrophysics Source Code Library, record
  ascl:2302.021

\bibitem[\protect\citeauthoryear{{Soulain} et~al.,}{{Soulain}
  et~al.}{2020}]{Soulain_2020}
{Soulain} A.,  et~al., 2020, in {Tuthill} P.~G.,  {M{\'e}rand} A.,   {Sallum}
  S.,  eds,  Society of Photo-Optical Instrumentation Engineers (SPIE)
  Conference Series Vol. 11446, Optical and Infrared Interferometry and Imaging
  VII. p. 1144611 (\mn@eprint {arXiv} {2201.01524}),
  \mn@doi{10.1117/12.2560804}

\bibitem[\protect\citeauthoryear{{St{\"u}rmer} \& {Quirrenbach}}{{St{\"u}rmer}
  \& {Quirrenbach}}{2012}]{Sturmer_2012}
{St{\"u}rmer} J.,  {Quirrenbach} A.,  2012, in {Delplancke} F.,  {Rajagopal}
  J.~K.,   {Malbet} F.,  eds,  Society of Photo-Optical Instrumentation
  Engineers (SPIE) Conference Series Vol. 8445, Optical and Infrared
  Interferometry III. p. 84452H, \mn@doi{10.1117/12.926270}

\bibitem[\protect\citeauthoryear{{Thatte}, {Greenbaum}, {McGruder},
  {Stansberry}  \& {Sivaramakrishnan}}{{Thatte}
  et~al.}{2016}]{2016LPI....47.3005T}
{Thatte} D.,  {Greenbaum} A.,  {McGruder} C.,  {Stansberry} J.,
  {Sivaramakrishnan} A.,  2016, in 47th Annual Lunar and Planetary Science
  Conference. Lunar and Planetary Science Conference.
p.~3005

\bibitem[\protect\citeauthoryear{{Tuthill}, {Monnier}  \& {Danchi}}{{Tuthill}
  et~al.}{1999}]{Tuthill_1999}
{Tuthill} P.~G.,  {Monnier} J.~D.,   {Danchi} W.~C.,  1999, \mn@doi [\nat]
  {10.1038/19033}, \href
  {https://ui.adsabs.harvard.edu/abs/1999Natur.398..487T} {398, 487}

\bibitem[\protect\citeauthoryear{Ulyanov, Vedaldi  \& Lempitsky}{Ulyanov
  et~al.}{2020}]{Ulyanov_2020}
Ulyanov D.,  Vedaldi A.,   Lempitsky V.,  2020, \mn@doi [International Journal
  of Computer Vision] {10.1007/s11263-020-01303-4}, 128, 1867

\bibitem[\protect\citeauthoryear{{Veeder}, {Davies}, {Matson}, {Johnson},
  {Williams}  \& {Radebaugh}}{{Veeder} et~al.}{2015}]{Veeder_2015}
{Veeder} G.~J.,  {Davies} A.~G.,  {Matson} D.~L.,  {Johnson} T.~V.,  {Williams}
  D.~A.,   {Radebaugh} J.,  2015, \mn@doi [\icarus]
  {10.1016/j.icarus.2014.07.028}, \href
  {https://ui.adsabs.harvard.edu/abs/2015Icar..245..379V} {245, 379}

\bibitem[\protect\citeauthoryear{Veverka, Helfenstein, Skypeck  \&
  Thomas}{Veverka et~al.}{1989}]{Veverka_1989}
Veverka J.,  Helfenstein P.,  Skypeck A.,   Thomas P.,  1989, \mn@doi [Icarus]
  {https://doi.org/10.1016/0019-1035(89)90066-3}, 78, 14

\bibitem[\protect\citeauthoryear{Wang \& Gang}{Wang \& Gang}{2018}]{Wang_2018}
Wang W.,  Gang J.,  2018, in 2018 International Conference on Information
  Systems and Computer Aided Education (ICISCAE). pp 64--70,
  \mn@doi{10.1109/ICISCAE.2018.8666928}

\bibitem[\protect\citeauthoryear{{Wizinowich}, {Acton}, {Lai}, {Gathright},
  {Lupton}  \& {Stomski}}{{Wizinowich} et~al.}{2000}]{wizinowich2000}
{Wizinowich} P.~L.,  {Acton} D.~S.,  {Lai} O.,  {Gathright} J.,  {Lupton} W.,
  {Stomski} P.~J.,  2000, in {Wizinowich} P.~L.,  ed.,  Society of
  Photo-Optical Instrumentation Engineers (SPIE) Conference Series Vol. 4007,
  Adaptive Optical Systems Technology. pp 2--13, \mn@doi{10.1117/12.390368}

\bibitem[\protect\citeauthoryear{Woods}{Woods}{2012}]{Woods_2012}
Woods J.~W.,  2012, in Woods J.~W.,  ed., , Multidimensional Signal, Image, and
  Video Processing and Coding (Second Edition), second edition edn, Academic
  Press, Boston, pp 223--256,
  \mn@doi{https://doi.org/10.1016/B978-0-12-381420-3.00007-2}, \url
  {https://www.sciencedirect.com/science/article/pii/B9780123814203000072}

\bibitem[\protect\citeauthoryear{Zhang \& Brandt}{Zhang \&
  Brandt}{2021}]{Zhang_2021}
Zhang H.,  Brandt T.~D.,  2021, \mn@doi [The Astronomical Journal]
  {10.3847/1538-3881/ac1348}, 162, 139

\bibitem[\protect\citeauthoryear{{de Kleer} \& {de Pater}}{{de Kleer} \& {de
  Pater}}{2016}]{de_Kleer_2016b}
{de Kleer} K.,  {de Pater} I.,  2016, \mn@doi [\icarus]
  {10.1016/j.icarus.2016.06.019}, \href
  {https://ui.adsabs.harvard.edu/abs/2016Icar..280..378D} {280, 378}

\bibitem[\protect\citeauthoryear{{de Kleer} et~al.,}{{de Kleer}
  et~al.}{2019}]{de_Kleer_2019}
{de Kleer} K.,  et~al., 2019, \mn@doi [\aj] {10.3847/1538-3881/ab2380}, \href
  {https://ui.adsabs.harvard.edu/abs/2019AJ....158...29D} {158, 29}

\bibitem[\protect\citeauthoryear{{de Kleer}, {Hughes}, {Nimmo}, {Eiler},
  {Hofmann}, {Luszcz-Cook}  \& {Mandt}}{{de Kleer} et~al.}{2024}]{dekleer2024}
{de Kleer} K.,  {Hughes} E.~C.,  {Nimmo} F.,  {Eiler} J.,  {Hofmann} A.~E.,
  {Luszcz-Cook} S.,   {Mandt} K.,  2024, \mn@doi [Science]
  {10.1126/science.adj0625}, \href
  {https://ui.adsabs.harvard.edu/abs/2024Sci...384..682D} {384, 682}

\bibitem[\protect\citeauthoryear{{de Pater}, {Davies}, {McGregor}, {Trujillo},
  {{\'A}d{\'a}mkovics}, {Veeder}, {Matson}  \& {Leone}}{{de Pater}
  et~al.}{2014a}]{depater2014Io}
{de Pater} I.,  {Davies} A.~G.,  {McGregor} A.,  {Trujillo} C.,
  {{\'A}d{\'a}mkovics} M.,  {Veeder} G.~J.,  {Matson} D.~L.,   {Leone} G.,
  2014a, \mn@doi [\icarus] {10.1016/j.icarus.2014.06.019}, \href
  {https://ui.adsabs.harvard.edu/abs/2014Icar..242..379D} {242, 379}

\bibitem[\protect\citeauthoryear{{de Pater}, Davies, McGregor, Trujillo,
  Ádámkovics, Veeder, Matson  \& Leone}{{de Pater}
  et~al.}{2014b}]{DEPATER_2014}
{de Pater} I.,  Davies A.~G.,  McGregor A.,  Trujillo C.,  Ádámkovics M.,
  Veeder G.~J.,  Matson D.~L.,   Leone G.,  2014b, \mn@doi [Icarus]
  {https://doi.org/10.1016/j.icarus.2014.06.019}, 242, 379

\bibitem[\protect\citeauthoryear{{de Pater}, {de Kleer}, {Molter}, {Alvarez}
  \& {Campbell}}{{de Pater} et~al.}{2017}]{depatertwi}
{de Pater} I.,  {de Kleer} K.,  {Molter} E.,  {Alvarez} C.,   {Campbell} R.,
  2017, {The Twilight Zone Project},
  https://www2.keck.hawaii.edu/inst/tda/TwilightZone.html

\bibitem[\protect\citeauthoryear{{de Pater}, {de Kleer}  \&
  {{\'A}d{\'a}mkovics}}{{de Pater} et~al.}{2020a}]{depater2020IoSO}
{de Pater} I.,  {de Kleer} K.,   {{\'A}d{\'a}mkovics} M.,  2020a, \mn@doi [The
  Planetary Science Journal] {10.3847/PSJ/ab9eb1}, \href
  {https://ui.adsabs.harvard.edu/abs/2020PSJ.....1...29D} {1, 29}

\bibitem[\protect\citeauthoryear{{de Pater}, {Luszcz-Cook}, {Rojo}, {Redwing},
  {de Kleer}  \& {Moullet}}{{de Pater} et~al.}{2020b}]{de_Pater_2020}
{de Pater} I.,  {Luszcz-Cook} S.,  {Rojo} P.,  {Redwing} E.,  {de Kleer} K.,
  {Moullet} A.,  2020b, \mn@doi [The Planetary Science Journal]
  {10.3847/PSJ/abb93d}, \href
  {https://ui.adsabs.harvard.edu/abs/2020PSJ.....1...60D} {1, 60}

\bibitem[\protect\citeauthoryear{{de Pater}, {Keane}, {de Kleer}  \&
  {Davies}}{{de Pater} et~al.}{2021}]{de_Pater_2021}
{de Pater} I.,  {Keane} J.~T.,  {de Kleer} K.,   {Davies} A.~G.,  2021, \mn@doi
  [Annual Review of Earth and Planetary Sciences]
  {10.1146/annurev-earth-082420-095244}, \href
  {https://ui.adsabs.harvard.edu/abs/2021AREPS..49..643D} {49}

\bibitem[\protect\citeauthoryear{{de Pater} et~al.,}{{de Pater}
  et~al.}{2022}]{depater2022JWSTDPS}
{de Pater} I.,  et~al., 2022, in AAS/Division for Planetary Sciences Meeting
  Abstracts. p. 306.07

\bibitem[\protect\citeauthoryear{{de Pater} et~al.,}{{de Pater}
  et~al.}{2023}]{de_Pater_2023}
{de Pater} I.,  et~al., 2023, \mn@doi [Journal of Geophysical Research
  (Planets)] {10.1029/2023JE007872}, \href
  {https://ui.adsabs.harvard.edu/abs/2023JGRE..12807872D} {128, e2023JE007872}

\makeatother
\end{thebibliography}




\appendix

\section{Glossary of Neural Network Components}
\label{sec:glossary}
In the following list, we provide small technical descriptions of the different components used by the neural networks presented in this work.

\begin{description}[align=left]

\item[1. Nodes/Neurons:] Interconnected components within a neural network, where each connection represents a weight where knowledge is stored through a learning process.
\item[2. Weights:] Internal variables within a neural network that are adjusted during training to minimize a loss function, enabling the network to learn complex, non-linear mappings from inputs to outputs.
\item[3. Loss Function] A function that quantifies the difference between the predicted and actual outputs, which the network aims to minimize during training through iterative optimization.
\item[4. Convolutional Neural Networks (CNNs):] A type of neural network well-suited for image processing due to their ability to learn hierarchical features from data. They operate by learning hierarchical features directly from the input images. This is achieved through the use of kernels that scan the input, producing activation maps that highlight  the relevant features that compose images.
\item[5. Dense Layer:] A Dense Layer, also referred to as a fully connected layer. Each neuron in a dense layer is connected to every neuron in the preceding layer, allowing it to learn complex patterns across all input features. This type of layer is used as the first (input) layer in the model described in Sec. \ref{sec:cnn}
\item[6. Convolutional Layer:] A common component in neural networks for image processing by applying kernels. These filters slide over the input, computing dot products between their weights and local regions of the input, which in turn generates an "activation map" or "feature map" that highlights specific patterns or features within the data. This type of layers is the core of the architectures presented in Sec. \ref{sec:cnn} and \ref{sec:dip}
\item[7. Learnable Filters (or Kernels):] A kernel is a small matrix of learnable weights. This kernel slides across the input data (e.g., an image or a feature map) to perform a convolution operation. The essential operation performed by a kernel can be described as a summation of element-wise products. For a 2D input image $I$ and a 2D kernel $K$, the output feature map $S$ at a specific position $(i, j)$ is calculated as:

$$S(i, j) = (I * K)(i, j) = \sum_{m=0}^{M-1} \sum_{n=0}^{N-1} I(i+m, j+n) \cdot K(m, n)\,,$$

where $I(x, y)$ represents the pixel value at position $(x, y)$ in the input image, $K(m, n)$ represents the weight of the kernel at position $(m, n)$, $M \times N$ is the size of the kernel. The summation is performed over the region of the input image covered by the kernel at position $(i, j)$.

Essentially, the kernel's values are multiplied element-by-element with the corresponding pixel values in the receptive field of the input, and all these products are summed up to produce a single output value for that specific location in the feature map. This process is repeated as the kernel slides across the entire input, generating the complete feature map.
\item[8. Activation Map (Feature Map):] A 2D map produced as a filter slides over the input, highlighting regions activated by the filter and capturing important features like edges or textures.
\item[9. Activation Function:] The essential operation performed by an activation function to a neuron is to introduce non-linearity into its output. After a neuron calculates a weighted sum of its inputs, the activation function processes this sum. This transformation determines whether the neuron should be activated and what value it should pass on as output to subsequent layers. 
\item[10. ReLU Activation Function:] The Rectified Linear Unit (ReLU) activation function introduces non-linearity by simply thresholding values at zero. If the input to the ReLU function is positive, it outputs the input value itself. If the input is negative, it outputs zero. This straightforward operation helps in mitigating the vanishing gradient problem, which can hinder the training of deep networks when using other activation functions like sigmoid or tanh. The equation for the ReLU activation function is given by:

$$f(x) = \max(0, x)\,,$$

where $x$ is the input to the activation function (the weighted sum of inputs to the neuron) and $f(x)$ is the output of the activation function.

\item[11. AReLU Activation Function:] The Adaptive Rectified Linear Unit (AReLU) activation function is a trainable activation function that provides a more flexible approach to introducing non-linearity compared to fixed functions like ReLU. Its adaptability comes from the inclusion of a trainable weight. It is defined as:

$$f(x) = \max(Ax, x)\,,$$

where $x$ is the input, $A$ is a trainable weight, meaning its value is adjusted during the training process of the neural network. $f(x)$ is the output of the AReLU activation function. This adaptive nature allows the network to learn the optimal non-linear transformation for its specific task, potentially leading to improved performance compared to using static activation functions.

\item[12. LeakyReLU:] The Leaky Rectified Linear Unit (Leaky ReLU) activation function is a variation of the ReLU activation function designed to address the dying ReLU problem. The standard ReLU outputs zero for all negative inputs, which can lead to neurons becoming inactive and subsequently ceasing to learn. Leaky ReLU mitigates this issue by allowing a small, non-zero gradient when the input is negative. Instead of outputting zero for negative inputs, it outputs a small linear component of the input. This ensures that neurons can still have a non-zero gradient even when their activation is negative, allowing for continued learning. The equation for the Leaky ReLU activation function is given by:

$$f(x) = \begin{cases} x & \text{if } x > 0 \\ \alpha x & \text{if } x \le 0 \end{cases}\,,$$

where $x$ is the input to the activation function (the weighted sum of inputs to the neuron). $\alpha$ (alpha) is a small, fixed positive constant (typically a small value like 0.01) and $f(x)$ is the output of the activation function. The LeakyReLU activation layer was used in the compression and decompression blocks of the unsupervised Deep Image Prior neural network.
\item[13. Autoencoder:] A type of unsupervised forward model that represents high-dimensional data in a lower-dimensional latent space from which a model of the original image is recovered. In DIP, it serves as a denoiser.
\item[14. Latent Space:] A lower-dimensional representation of high-dimensional data, used by autoencoders to recover a model of the original image.
\item[15. Skip Connections:] Skip connections are a critical architectural feature used in neural networks, particularly in U-Net-like architectures such as the Deep Image Prior DIP neural network described in Sec. \ref{sec:dip}. They serve to address a limitation inherent in deep convolutional networks, specifically the challenge of learning correlations between non-adjacent pixels. In a typical autoencoder the data undergoes significant downsampling in the encoder path, where spatial information can be lost. When the network then attempts to reconstruct the output in the decoder path through upsampling, fine-grained details might not be accurately recovered due to this loss of information. Skip connections work by directly connecting layers from the encoder path to corresponding layers in the decoder path. This allows the network to bypass some layers in the processing pipeline and directly pass information from earlier, higher-resolution layers to later, lower-resolution layers. By doing so, skip connections preserve fine-grained details, improve gradient flow and facilitate learning.
\item[16. Batch Normalization Layer:] A Batch Normalization layer is a technique used in training deep neural networks to normalize the inputs of each layer. This normalization is performed by re-centering and re-scaling the inputs to have a mean of zero and a standard deviation of one. The operation is typically applied to mini-batches of data during training. The batch Normalization layer was a component within both the compression and decompression blocks of the DIP network.
\item[17. Reflection Padding 2D Layer:] This is a type of padding technique used in convolutional neural networks. Its primary purpose is to enlarge the spatial dimensions of an input tensor (e.g., an image or feature map) before it undergoes convolutional operations.
\item[18. Adam Optimizer:] The Adaptive Moment Estimation (Adam) optimizer is an adaptive learning rate optimization algorithm that can handle sparse gradients on noisy problems.
\item[19. Noise-Based Regularization:] This is a family of techniques used in training neural networks to prevent overfitting and improve the model's generalization capabilities. The core idea is to introduce a controlled amount of random noise into different parts of the network during the training phase. This makes the model more robust and less sensitive to minor fluctuations in the input data, effectively smoothing the optimization landscape and encouraging the network to learn more generalized features rather than memorizing the training data, including its noise. This technique was used in the DIP network and it  adds normal noise with zero mean and a standard deviation $\sigma_{p}=1/30$ to the input signal.

\end{description}

\section{Simulations for flux retrieval characterization}
\label{sec:disk_sim_flux}
To investigate the effect of reporting just the radiance at the brightest pixel for each hot spot on the surface of Io, a series of numerical simulations were conducted. These simulations aimed to emulate potential substructures, akin to the sources  V1-V4 observed in the DIP images. A synthetic model was constructed comprising a uniform circular disk with the embedded presence of four distinct bright spots. The pixel scale of this simulation was set to be ten times finer than the native pixel scale of the JWST-NIRISS camera. This oversampling allowed for a more precise characterization of the sub-pixel positioning of the hot spots, prior to convolution and subsequent rescaling. The total integrated emission from the simulated disk and bright spots was normalized to unity. The diffuse emission from the disk was assigned 90\% of the total flux, while the remaining 10\% was equally distributed among the four embedded bright spots. Each of the four bright spots was convolved with a two-dimensional Gaussian kernel representing the synthesized beam of the AMI data. The full-width at one-tenth (0.10) maximum (FWTM) of this Gaussian is set to $\lambda / 2B = 93.18$ mas, where $\lambda$ = 4.3 $\mu$m and B = 4.76 m. 

A Monte Carlo approach was implemented to assess the impact of positional variations on flux measurements. For each of the four bright spots, their initial coordinates were randomly perturbed within a uniform distribution ranging from -5 to +5 pixels in both R.A. and Dec. Given the initial oversampling factor of 10, this perturbation range corresponds to a spatial uncertainty of $\pm 0.5$ pixels at the JWST scale plate. A total of 1000 independent realizations of the disk with these randomly perturbed bright spot positions were generated to form a robust statistical ensemble.

Each of the simulated images was rescaled to the native pixel resolution of JWST-NIRISS. For each realization, the maximum pixel value of each bright spot was recorded. Subsequently, the average peak flux across all 1000 simulations for each bright spot was calculated, along with the corresponding standard deviation. This analysis revealed that, on average, the flux of the brightest pixel associated with a bright spot constituted $40 \pm 5\%$ of the total integrated emission of that specific bright spot.

\begin{figure}
    \includegraphics[width = 7 cm]{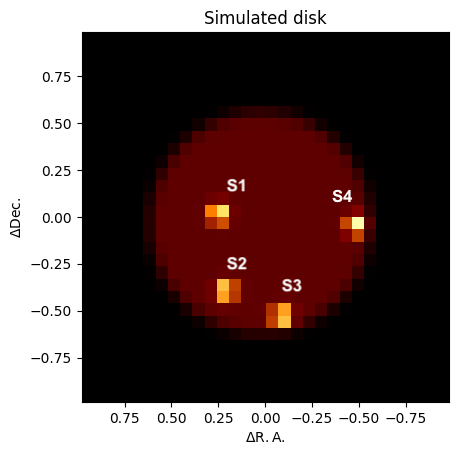}
    \caption{Example of the simulated disk + bright spots for the flux retrieval characterization experiment. The image is at the resolution of the JWST-NIRISS instument (65.7 mas px$^{-1}$) The four bright spots simulated are labeled S1-S4. The initial positions of the simulated bright spots were obtained from the positions of V1-V4 of the DIP images for the first AMI epoch (2022-08-01, 16:32:23 UT).}
    \label{fig:sim_disk}
\end{figure}

In addition to the peak flux measurements, the integrated emission within a $3 \times 3$ pixel aperture centered on the peak emission of each bright spot was computed for every simulation. This aperture size was chosen to encompass the majority of the flux associated with the convolved bright spots while minimizing the contribution from the surrounding diffuse disk emission. By averaging the integrated flux values across the 1000 simulations, we quantified the fraction of the initial bright spot flux that could be recovered through aperture photometry. For the two bright spots (S1 and S2 in Fig.\,\ref{fig:sim_disk}) situated well within the radial extent of the disk, the average recovered integrated flux was $100 \pm 1\%$ of their initial contribution. However, for the two bright spots positioned closer to the edge of the disk, the recovered integrated fluxes were $70 \pm 11\%$ and $80 \pm 6\%$ for S3 and S4, respectively. This indicates a potential loss of flux. We have to highlight that the simple uniform disk model used here does not account for more complex changes in the illumination of the disk and additional extended emission that could be present in the moon's surface, so it is not rate that the flux retrieval of the volcanoes at the edge of the disk suffers of increased contamination from the background disk emission for features located at larger radii.

\begin{figure*}
    \includegraphics[width = 16 cm]{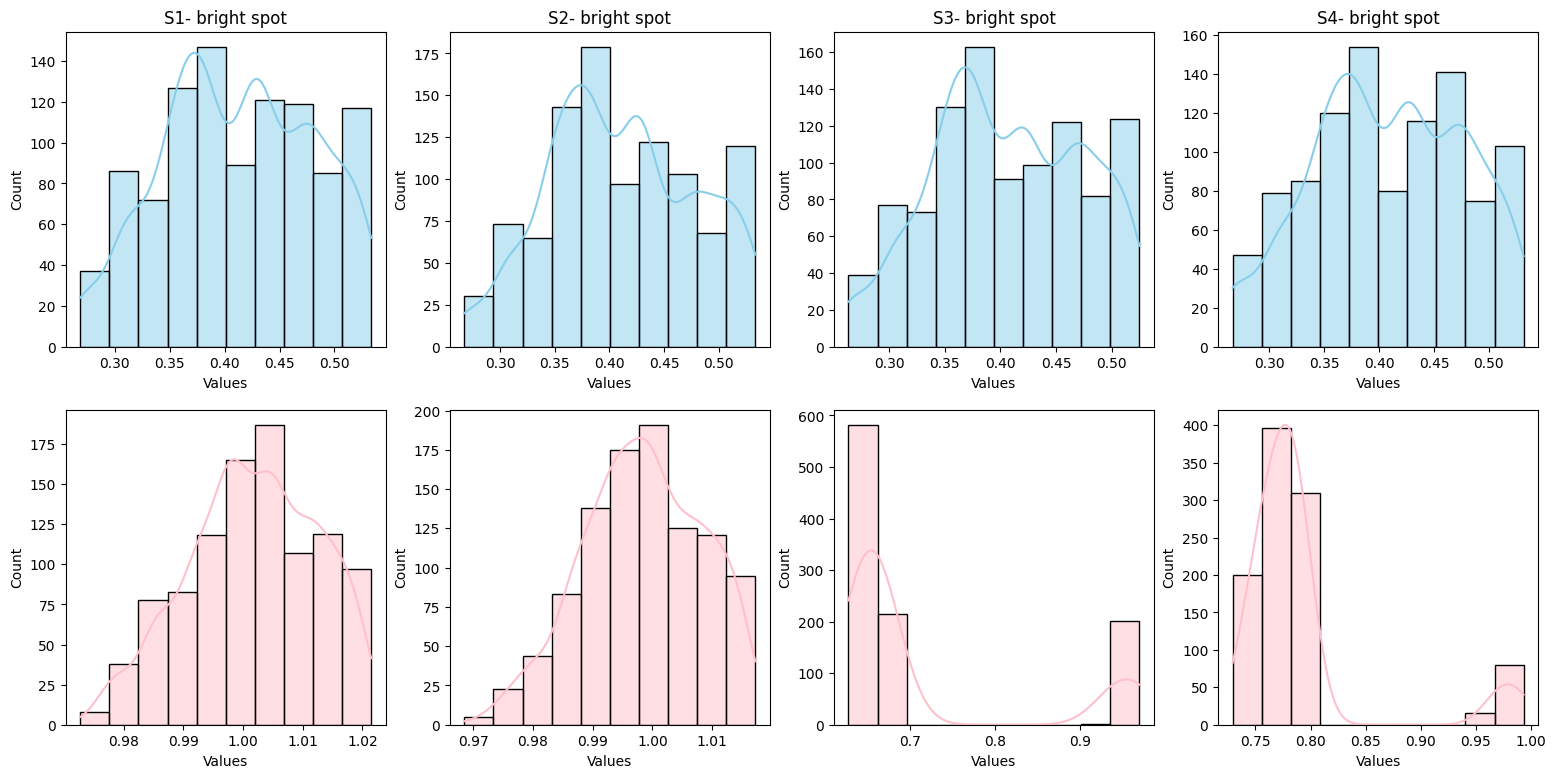}
    \caption{The upper row displays the histograms of the simulated bright spots S1-S4, considering only the peak flux. The lower row displays the corresponding histograms of S1-S4 considering the 3 $\times$ 3 pixel aperture centered at the position of their peak emission. Notice that with the aperture method for most of the samples in S3 and S4 recover between 70 to 80 \% of the total emission, and only in a few realizations the total emission is recovered }
    \label{fig:sim_disk}
\end{figure*}

\section{Projection of Io disk in a 2D longitude-latitude map}

\begin{figure*}
    \includegraphics[width=14 cm]{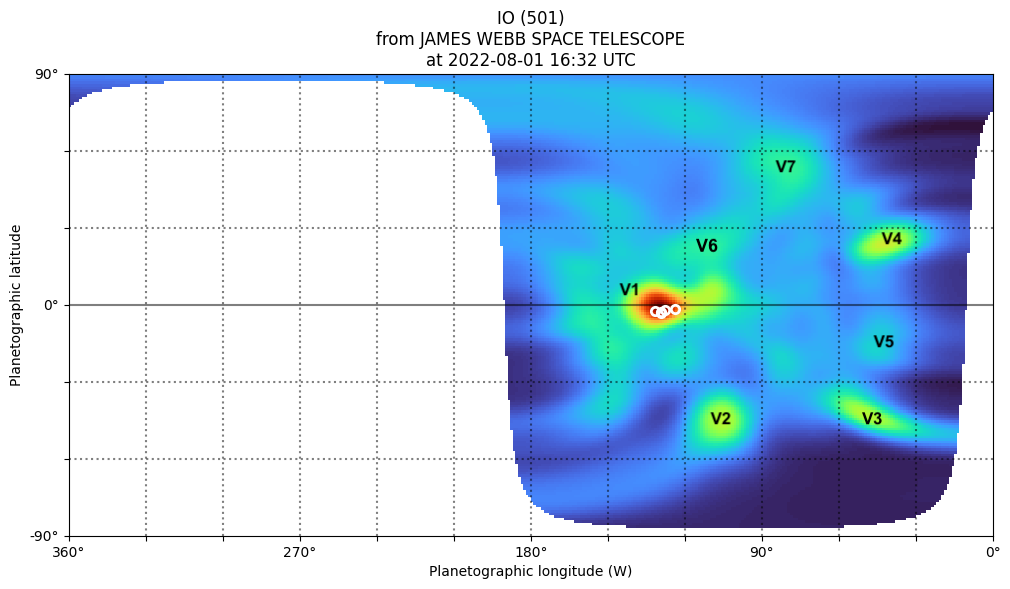}
    \caption{Orthographic projection of Io's disk on a Longitude - Latitude map. The image displays the projected DIP deconvolved disk at the ephemeris of the first epoch (2022-08-01 16:32:23) of our science exposures. The sampling of the map is 1 deg px$^{-1}$, the original DIP images were reescaled at this resolution using a cubic interpolation. The seven volcanoes identified are labeled in the panel. The white unfilled dots show the mean flux centroid positions of V1 for the five different epochs in our AMI data. Notice the displacement of the centroids due to Io's rotation. }
    \label{fig:projection_io}
\end{figure*}



\bsp	
\label{lastpage}
\end{document}